\documentclass[journal=jacsat,manuscript=article]{achemso}

\usepackage{graphicx}
\usepackage{physics}
\usepackage{braket}
\usepackage{mathrsfs}
\usepackage{placeins}
\usepackage{booktabs, tabularx} % added_AS for table in text
\usepackage{multirow}
\usepackage{color}
\usepackage[normalem]{ulem}
\usepackage[hidelinks]{hyperref} % AS Added for ref hyperlink and removing the box around 
\usepackage[version=3]{mhchem} % Formula subscripts using \ce{}

\author{Atandrita Bhattacharyya}
\author{Amitav Sahu}
\author{Sanjoy Patra}
\author{Vivek Tiwari}
\email{vivektiwari@iisc.ac.in}
\affiliation{Solid State and Structural Chemistry Unit, Indian Institute of Science, Bangalore, Karnataka 560012, India}

\renewcommand{\sout}[1]{\unskip}
\newcolumntype{C}{>{\centering\arraybackslash}X}% added_AS for table in text
\SectionNumbersOn

\newcommand*{\vt}[1]{\textcolor{black}{ #1}}
\newcommand*{\si}[1]{\textcolor{black}{ #1}}

\title {Low and High Frequency Vibrations Synergistically Enhance Singlet Exciton Fission Through Robust Vibronic Resonances}

\begin{document}

\begin{abstract}
	Singlet exciton fission (SEF) is initiated by ultrafast internal conversion of a singlet exciton into a correlated triplet pair $(TT)^1$. The `reaction coordinates' for ultrafast SEF even in archetypal systems such as pentacene thin film remain unclear with synthetic design principles broadly relying on tailoring electronic couplings to achieve new templates for efficient SEF materials.  Spectroscopic detection of vibrational coherences in the $(TT)^1$ photoproduct has motivated theoretical investigations into a possible role of vibronic resonance in driving SEF, akin to that reported in several photosynthetic proteins. However, a precise understanding of how prominent low-frequency vibrations and their modulation of intermolecular orbital overlaps, equally prominent high-frequency vibrations, and order of magnitude larger Huang-Rhys factors in SEF chromophores compared to photosynthetic pigments, collectively influence the mechanistic details of SEF remains starkly lacking. Here we address this gap and identify previously unrecognized effects which are quite contrasting from those known in photosynthesis excitons, and vitally enhance non-adiabatic internal conversion in SEF. These include broad non-selective vibronic resonances, high-frequency vibrations coupling multiple such resonances, and a synergistic effect of low-frequency vibrations on vibronic resonances. Spectroscopically these effects are reported by quantum beats of mixed polarization signatures in the simulated two-dimensional electronic spectra. We leverage this to propose readily implementable polarization-based experiments to test the above predictions, and unambiguously distinguish vibrations which drive vibronic mixing and promote SEF, against spectator vibrations simply accompanying ultrafast internal conversion. Our findings have direct implications for the broad experimental interest in synthetically tailoring molecules to promote vibronically enhanced internal conversion.  
\end{abstract}

\section{Introduction} 
%\sout{\ab{1. Check JACS format, for example, is it Figure 1A or Figure 1A ? What is the format for SI ? How is the SI referred to, Figure S1 of the SI or simply Figure S, etc. ?}}
%\sout{\ab{2.calculation with 150 cm-1 or 265 cm-1 frequency with no FC displacement, etc. supporting calcs summarize all these in a SI section and as a Table perhaps}}
%\vt{3. VT will do - a comment about Engel2021\cite{Engel2021} in intro ?}\\
%\sout{\ab{4. role of HT coupling in beta mixing and without low freq FC activity how is the P(t) changing, table for both.}}
%\ab{5. [from AB] Incorporating VT comments about coherence transfer (?)} \\
%\sout{\ab{6. Justification of c = 0.554.}}
%\vt{7. from VT - Justification of Ehrenfest Dynamics} \\
%\sout{\ab{[last] one long grid simulation to check if cBands are not arising due to early truncation}}\\

Ultrafast conversion of a singlet exciton into a correlated triplet pair $(TT)^1$ is the first step in singlet exciton fission (SEF)\cite{Michl2013}, and subsequently leads to the generation of spatially separated triplets with uncorrelated spin wavefunctions on slower timescales. The high efficiency of the overall process in certain materials promises applications in next-generation photovoltaic technologies. However, mechanistic details of even the first step of this reaction remain unsettled ranging from preliminary suggestions of direct\cite{Zimmerman2011}, charge transfer (CT) state mediated\cite{Ananth2014,Krylov2013} and `superexchange' mediated\cite{Berkelbach2} ultrafast conversion of $S_1$ to $(TT)^1$ to a more recently proposed role for increased charge-transfer couplings at crystal edges and defects\cite{Jones2020}. Molecular design principles to achieve new SEF chromophores have accordingly been generally limited to synthetic tuning of electronic couplings only. A role for mixed vibrational-electronic (vibronic) motions in driving this process has come under close scrutiny\cite{Tempelaar2017_1,Tempelaar2017_2,Tempelaar2018_3,Tempelaar2022,Herbert2017,Duan2020,Rao2016} only recently.  

In photosynthesis, likely coincidences between large number of Franck-Condon (FC) active low-frequency intramolecular vibrations of the pigments with excited state electronic energy gaps of the multi-pigment protein are termed as vibronic resonances\cite{Womick2011,Tiwari2013}. These lead to strongly coupled vibrational-electronic motions creating non-adiabatic energy funnels\cite{Peters2017} which can efficiently drive photosynthetic energy and charge transfer. Such resonances are increasingly reported in several photosynthetic proteins\cite{JonasARPC2018} based on readily observed vibronic coherences, or impulsively excited superpositions of vibronically mixed eigenstates. Along similar lines, and as a departure from previously proposed SEF mechanisms, vibronic resonances along high-frequency intramolecular vibrations in SEF chromophores have been recently implicated\cite{Rao2016, Tempelaar2018_3,Tempelaar2022} to drive ultrafast SEF. While this may be material specific, even for archetypal systems such as pentacene thin films it is not clear whether vibronic resonance due to a high-frequency intramolecular vibration is robust enough to energetic mismatches\cite{Tempelaar2018_3} to enhance the rate of $(TT)^1$ formation. A recent two-state \textit{ad hoc} model\cite{Duan2020} also suggests no role for vibronic resonance. While it is not clear whether models with two electronic states, ${S}_1$ and ${(TT)}^1$, adequately capture the essential physics of the first steps of SEF and will be a subject of this presentation, some pertinent questions regarding the physical principles governing the initial steps of SEF may be -- 1. Huang-Rhys (HR) factors, which dictate the probability of vibrational excitations accompanying electronic excitation, are order of magnitude larger in acenes than in photosynthetic pigments. How does that modify the vibronic resonances in SEF compared to narrow, selective vibronic resonances\cite{Tiwari2017, Tiwari2018} between photosynthetic excitons ? 2. Prominent low-frequency FC active vibrations, which also tune intermolecular orbital overlaps are known in acene thin films\cite{Venuti2002,Venuti2004,Rao2016,Duan2020}. How do these vibrations of mixed inter/intramolecular character affect resonant vibronic couplings present along the high-frequency intramolecular vibrations ? 3. Large  HR factors in acenes imply that prominent vibrational quantum beats are expected even in a solution of monomers. Can one then decipher mechanistic details solely based on readily observed quantum beats ?  \\

% Joo and co-workers have recently illustrated\cite{Joo2020} that generation of vibrational wavepakets with \textit{new} coherence frequencies and their beating amplitude ultimately depends on the projection of relative FC displacements of reactant and product states on the normal modes of the photoproduct. In the related context of ultrafast SEF driven by and/or generating vibrational coherences,

The latter question deserves further motivation. In the archetypal SEF chromophores such as pentacene, large HR factors, with dimensionless displacements $d$ of the order of $d = \pm1$ (classical turning points of the zero-point level), imply that ultrafast population transients in time-resolved spectroscopies will also exhibit strong modulations due to oscillating vibrational wavepackets on the ground and excited electronic states, even for pentacene monomers in solution\cite{Tan2021}. Further complicating the interpretation of quantum beats in experiments is the well-known fact that dominant vibrational wavepackets in the photoproduct can simply arise due to transfer of coherently prepared reactant vibrational wavepackets that survive non-adiabatic internal conversion, or due to photoproduct formation within a fraction of vibrational time period\cite{Champion1994,Wynne1996,Zinth1998, Joo2012,Scholes2021}. For instance, Petelenz and co-workers have elucidated\cite{Petelenz2019,PetelenzJPCC2020} that experimentally reported\cite{Stern2017} new vibrational coherences in the $(TT)^1$ product may be those that are simply transferred from $S_1$ to $(TT)^1$ with substantially larger amplitude in the photoproduct. Multilevel Redfield simulations\cite{Jean1995} of electronic curve crossing from Jean and Fleming further emphasize the above points. They show that vibrational wavepackets along spectator vibrational modes orthogonal to the reaction coordinate can be coherently transferred to the photoproduct without participating in ultrafast internal conversion. In contrast, wavepacket motions along promoter modes that drive vibronic mixing experience significant anharmonicity and dephase during ultrafast internal conversion. The conclusion from above experimental and theoretical reports is that vibrational coherences or new vibrational frequencies in the photoproduct do not \textit{a priori} identify `reaction coordinates' for internal conversion. This is true even for state-of-the-art two-dimensional electronic spectroscopy (2DES) experiments, which resolve ultrafast dynamics along multiple spectral dimensions -- excitation, detection and coherence frequencies. For example, strong modulations reported in 2DES studies\cite{Rao2016,Duan2020} on pentacene thin films may simply arise from spectator vibrational wavepackets which are not dephased by non-adiabatic anharmonic couplings present along promoter modes, and continue to oscillate through the SEF process. Thus, while much interest is generated by the possibility of selectively exciting promoter modes\cite{Paulus2020,Weinstein2014} and leveraging vibronic couplings\cite{Scholes2017} to select reaction outcomes, unique spectroscopic markers of promoter modes are vital to decipher the quantum dynamical details of ultrafast SEF. \\
%Here we will address the above interpretational ambiguities regarding the experimentally observed quantum beats to propose polarization-based 2DES experiments for unambiguously identifying the promoter modes in this process. \\

This manuscript addresses the above questions, identifies previously unrecognized synergistic effects between high- and low-frequency vibrational motions which together enhance non-adiabatic internal conversion, and proposes readily implementable spectroscopic signatures that only arise from coupled vibrational-electronic motions without any ambiguity from spectator vibrations which do not promote SEF and simply accompany ultrafast internal conversion. Our findings have direct implications for synthetic design of molecules to promote vibronically enhanced SEF, as well as for spectroscopy experiments aiming to distill the relevant mechanistic details.\\

\section{Results and Discussion}

Analysis of vibronic coupling in a donor-acceptor system provides a useful starting point to distinguish promoter versus spectator vibrational motions. For example, the Hamiltonian $\hat{H}(\hat{q}_A,\hat{q}_B)$ of an electronically coupled dimer can be analyzed in terms of in-phase and anti-phase relative motions of the two molecules $A$ and $B$ along their respective intramolecular vibrational coordinates $\hat{q}_{A,B}$. These can be represented as $\hat{q}_{\pm}$, formed by linear combinations of two intramolecular Franck-Condon (FC) modes on the respective molecules. This has precedence in the early works of Witkowskii and Moffit\cite{Moffit1960}, Fulton and Gouterman\cite{Gouterman1961}, and F\"{o}rster\cite{Sinanoglu}, and leads to adiabatic separability of in-phase motions along $\hat{q}_+$ from the rest of the Hamiltonian. In-phase motions do not affect donor-acceptor energy gap and are mere spectators in the dimer energy transfer problem, while anti-phase motions along $\hat{q}_-$ \textit{tune} relative donor-acceptor energy gap and act as promoter modes that mix electronic degrees of freedom.  At vibronic resonance, this mixing becomes strongly non-adiabatic\cite{Tiwari2013, Peters2017} and only anti-phase motions drive the vibronic wavepacket towards the acceptor. Here we adopt this understanding to analyze the SEF Hamiltonian.
%Jonas and co-workers have shown that the loss of anti-phase relationship also naturally describes decoherence\cite{Tiwari2017,Tiwari2018} between excited states. 

Berkelbach and Reichman\cite{Berkelbach2,Berkelbach3} have described ultrafast SEF in a pentacene dimer unit cell as well as supercells using a microsopic exciton model. Their calculations suggest that a dimer description can still qualitatively describe SEF rates. In larger systems\cite{Persico2020,Berkelbach3} with growing CT manifold, or with stabilized\cite{Nakano2021} CT states in the crystal, increased CT-mediated mixing of $S_1$ and $(TT)^1$ further enhances SEF rates. \textit{Ab initio} electronic structure calculations from Martinez \cite{Martinez2020} et al. have reported similar effects in going from dimer to bulk crystals. Zimmerman et al. have reported\cite{Zimmerman2011} that the initial $S_1$ excitation at the vertical FC geometry is primarily dimeric, with contributions from neighboring pentacene molecules becoming negligible as the intermolecular distance is reduced by a small fraction ($\sim$20\%) post photoexcitation. Overall, a dimer unit cell description for the archetypal pentacene system is a reasonable starting point. A further entropic enhancement of SEF rates due to a large number of final $(TT)^1$ manifold of states, as suggested by Tempelaar et al.\cite{Tempelaar2018_3}, will not be captured in the dimer description adopted here.  
%JACS{incorporating quantum relaxation through a Markovian Redfield bath. 
%to illustrate previously unrecognized effects which can enhance SEF such as broad, non-selective and coupled vibronic %resonances, and the synergistic interplay of low and high frequency vibrations.}
%Tempelaar et al. have remarked that, in case of supercells with only $N$ singlet states which are further reduced to only the lower Davydov component having substantial oscillator strength, the $N(N-1)/2$ triplet manifold leads to thermodynamically expected faster SEF rates. Such an 
\begin{figure*}[ht]
	\centering	\includegraphics[width=2.5 in] {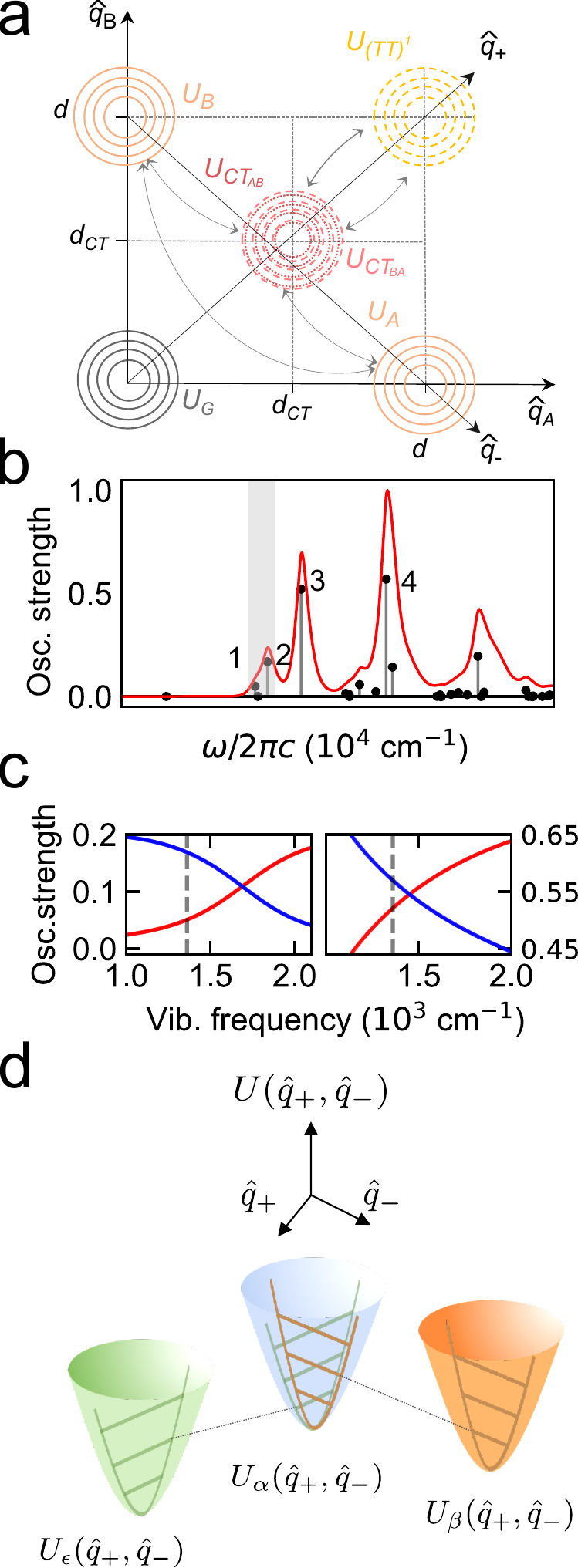}
	\caption{\footnotesize (\textbf{a}) Schematic shows two dimensional contours of diabatic site basis potentials as a function of intramolecular vibrational coordinates, $\hat{q}_A$ and $\hat{q}_B$, corresponding to one high frequency intramolecular vibration on either molecule. The diabatic potentials are denoted by $U_S(\hat{q}_A ,\hat{q}_B)$ where $S \in \{G,A,B,CT_{AB},CT_{BA},(TT)^1\}$. Excluding the ground state, optically bright (dark) basis states are denoted by solid (dashed) contours. $CT_{BA}$ is deliberately shown with dotted contours for the sake of clarity. Arrows denote electronic couplings between the diabatic states. (\textbf{b}) Linear absorption spectrum at 300 K for the case of one explicit high-frequency vibrational mode ($\omega_H = 1360$ cm$^{-1}$) on each molecule in the system Hamiltonian. `Sticks' denote oscillator strengths arising from the system Hamiltonian, while lineshapes arise when the Brownian oscillator bath is also included. The faint grey band denotes the (10)$\epsilon\alpha$ manifold. Details of the calculation are described in \textcolor{black}{Section S3}. (\textbf{c}) Intensity borrowing between marked pairs of peaks (left panel for peaks 1,2, right panel for peaks 3,4) plotted as a function of vibrational frequency while keeping $\hat{H}_{elec}$ fixed. The vertical dashed line correspond to $\omega_H = 1360$ cm$^{-1}$. (\textbf{d}) Coupled $\epsilon-\alpha$ and $\alpha-\beta$ resonances between diabatic excitonic potentials corresponding to $\epsilon$, $\alpha$ and $\beta$ excitons, along $\hat{q}_+$ and $\hat{q}_-$ vibrational modes.}
	
	\label{fig:fig1}
\end{figure*}
\FloatBarrier
%\ab{Can resonance width for panel c be also derived using population transfer similar to figure 3c for an equivalent comparison, check this? Check SI.}
\subsection{Model Hamiltonian and Non-Trivial Role of In-phase Vibrational Motions}\label{qplus}
Following previous works\cite{Grozema2014,Herbert2017,Berkelbach1,Tempelaar2017_1}, the diabatic basis set for describing the singly excited manifold of coupled identical pentacene molecules $A$ and $B$  is given by  locally excited states $\ket{A}$ and $\ket{B}$, correlated triplet state $\ket{TT}$ and CT states $\ket{CT_{AB}}$ and $\ket{CT_{BA}}$. The notation adopted is such that $\ket{A}$ denotes a collective state where molecule $A$ is excited while molecule $B$ is in the ground electronic state, and $\ket{CT_{AB}}$ denotes a state in which $A$ is cationic and $B$ is anionic. The ground to excited state transition dipoles in pentacene are aligned along the short axis\cite{Martinez2020,Tempelaar2017_1} such that the transition dipoles to the optically allowed $\ket{A},\ket{B}$ states are $\sim$54.3$^o$ in the herringbone structure\cite{Hestand2015}. The doubly excited state manifold, $\ket{AB}$ will be discussed during the discussion of 2DES simulations. The purely electronic part of the singly excited Hamiltonian is given by $\hat{H}_{elec} = \sum_{L} \big[\epsilon_L \ket{L}\bra{L} + \sum_{K<L} J_{KL}(\ket{K}\bra{L} + \ket{L}\bra{K})\big]$. Matrix elements $J_{KL}$ arise due to static Coulomb or charge transfer couplings and $\epsilon_L$ is the electronic site energy of state $\ket{L}$. Indices run over the diabatic basis states. We next introduce linear vibronic coupling arising from the set of $V$ intramolecular FC active modes on each molecule, with vibrational coordinate operators $\hat{q}_{A_j}$ and $\hat{q}_{B_j}$ for the $j^{th}$ set. The linear vibronic coupling terms are $-\omega_j d_j\hat{q}_{L_j}$ on $\ket{L}$ where $\ket{L} = \ket{A},\ket{B}$,  $-(\omega_j d_j^{+(-)}\hat{q}_{A_j} + \omega_j d_j^{-(+)}\hat{q}_{B_j})$ on $\ket{CT_{AB(BA)}}$, and  $-\omega_j d_j^{TT}(\hat{q}_{A_j} + \hat{q}_{B_j})$ on $\ket{TT}$. The cationic and anionic stabilization energies for pentacene vibrations in the high frequency region are reported to be approximately equal\cite{Bredas2007}. Thus, for simplicity, we assume equal cationic and anionic FC displacements, that is, $d_j^+ = d_j^- = d^{CT}_j$. Similarly, we have assumed equal\cite{Tempelaar2017_1,Tempelaar2017_2} FC displacements on the triplet states of molecules $A$ and $B$. The singly excited state SEF Hamiltonian $\hat{H}$ becomes --
\begin{eqnarray} 
	\hat{H} = \hat{H}_{elec} &+& \sum_{L=A,B}\big[\hat{H}_G - \sum_{j}^{V}\omega_{j}d_{j}\hat{q}_{L_j}\big]\ket{L}\bra{L} \nonumber \\  &+& \sum_{CT}\big[\hat{H}_G - \sum_{j}^{V}\omega_{j}d_{j}^{CT}(\hat{q}_{A_j}+\hat{q}_{B_j})\big]\ket{CT}\bra{CT} \nonumber \\  &+& \big[\hat{H}_G - \sum_{j=1}^{V}\omega_{j}d_{j}^{TT}(\hat{q}_{A_j}+\hat{q}_{B_j})\big]\ket{TT}\bra{TT},
	\label{eq1}
\end{eqnarray}
%JACS {The latter two linear vibronic coupling terms indicate a crucial distinction of the SEF Hamiltonian compared to an excitonically coupled dimer. The slight difference of FC displacements between cationic and anionic states, or between locally excited and triplet states, will manifest on a timescale much longer than the faster timescales of interest here.} 
where $\hat{H}_G$ is the ground state Hamiltonian given by $\hat{H}_{G} = \sum_{j}^{V}[{\frac{1}{2}\omega_{j}{(\hat{p}_{A_{j}}^2+\hat{q}_{A_{j}}^2)}} + {\frac{1}{2}\omega_{j}{(\hat{p}_{B_{j}}^2+\hat{q}_{B_{j}}^2)}}]$. Equation~\ref{eq1} assumes identical but displaced harmonic oscillators to describe the diabatic electronic potentials. It is important to note that ground state vibrational excitations are not restricted. Vibrational excitations are allowed on electronically unexcited molecule as well because such a molecule can still be part of an overall excited state in which the other molecule is electronically excited. Imposing such a restriction is equivalent to one-particle\cite{Philpott1969} or coherent scattering approximation\cite{Briggs2008}. We\cite{Sahu2020,Patra2022} and others\cite{Tiwari2018,Valkunas2014} have previously shown that such an approximation severely underestimates the scope of vibronic resonance, although several studies\cite{Rao2016,Duan2020} describing vibronic couplings in SEF have the made one-particle approximation (1PA). Low-frequency vibrations in pentacene thin films are FC active\cite{Venuti2002,Filipini1984,Venuti2004}, have mixed inter/intramolecular character, and modulate intermolecular orbital overlaps to affect electronic couplings\cite{Troisi2006}. This latter effect is akin to Herzberg-Teller type coupling\cite{Mancal2014,Mancal2018} and is termed as Peierls coupling\cite{Tempelaar2018_3,Duan2020} in the context of SEF. The interplay of Peierls coupling and linear vibronic coupling will be analyzed in Section ~\ref{peierls}.\\

 Following earlier works \cite{Moffit1960,Gouterman1961,Sinanoglu} on energy transfer in excitonic dimers, we transform Equation \ref{eq1} with $\hat{q}_{j\pm} = (\hat{q}_{Aj} \pm \hat{q}_{Bj})/{\sqrt{2}}$, the in-phase and anti-phase relative motions of the two molecules, -- 
\begin{eqnarray} 
	\hat{H} = \hat{H}_{elec} &+& \big[\hat{H}_G - \sum_{j}^{V}(\frac{\omega_{j}d_{j}\hat{q}_{j_+}}{\sqrt{2}}+\frac{\omega_{j}d_{j}\hat{q}_{j_-}}{\sqrt{2}})\big]\ket{A}\bra{A} \nonumber \\ &+&\big[\hat{H}_G - \sum_{j}^{V}(\frac{\omega_{j}d_{j}\hat{q}_{j_+}}{\sqrt{2}}-\frac{\omega_{j}d_{j}\hat{q}_{j_-}}{\sqrt{2}})\big]\ket{B}\bra{B}\nonumber \\ &+&\sum_{CT}\big[\hat{H}_G -  \sum_{j=1}^{V} \sqrt{2}\omega_{j}d_{j}^{CT}\hat{q}_{j_+}\big]\ket{CT}\bra{CT} \nonumber \\ &+&\big[\hat{H}_G - \sum_{j=1}^{V} \sqrt{2}\omega_{j}d_{j}^{TT}\hat{q}_{j_+}\big]\ket{TT}\bra{TT}.
	\label{eq2}
\end{eqnarray}
%Therefore, in the rotated diabatic excitonic basis\cite{Tiwari2018,Peters2017}, electronically off-diagonal linear vibronic couplings between $\ket{TT}$ and $\ket{A,B}$ electronic domains, will arise from $\hat{q}_+$ motions as well,
%$\hat{q}_{j_\pm}$ corresponds to correlated and anti-correlated vibrational motions between molecules $A$ and $B$, respectively. For excitonic dimers, it is well understood that correlated vibrational motions $\hat{q}_+$ are adiabatically separable and play no role in promoting vibronic mixing. 
What becomes immediately evident from Equation~\ref{eq2} is that not just $\hat{q}_-$, but also $\hat{q}_+$ vibrational motions are adiabatically inseparable from the SEF Hamiltonian. Given the intuition from a purely excitonic dimer, mixing between the optically bright excitons, admixtures of $\ket{A}$ and $\ket{B}$, by anti-phase vibrational motions which tune their relative energy gap is not surprising. However, counter intuitive to this established picture,  in-phase motions, which were mere spectators in the case of excitonic dimer, can now \textit{drive} the coupling between bright excitons and the $\ket{TT}$ state. \\

The resultant physical picture is motivated in Figure~\ref{fig:fig1}a. Further mathematical analysis of linear vibronic couplings in Equation \ref{eq2} is presented in \textcolor{black}{Equations S1--S2 and Figure S1}. The multidimensional diabatic potentials of Equation~\ref{eq1} are two-dimensional for the simplest case of one FC active intramolecular vibrational mode on each molecule, and denoted as contours along intramolecular vibrational coordinates, $\hat{q}_A$ and $\hat{q}_B$. The diabatic potential $U_{A(B)}$ for the locally excited states $\ket{A}(\ket{B})$, are shifted by FC displacement $d$ along the respective intramolecular vibrational coordinate. Akin to what is expected in an excitonic dimer\cite{Tiwari2017}, the vector joining the minima of $U_B$ and $U_A$, is the anti-phase energy gap tuning mode $\hat{q}_-$ which drives non-adiabatic vibronic mixing between $\ket{A}$ and $\ket{B}$, while $\hat{q}_+$ is orthogonal and plays no role in the process. In the SEF problem, however, a striking contrast is seen when considering $U_{(TT)^1}$ and $U_{A(B)}$ diabatic states. Now, both in-phase and anti-phase motions, along $\hat{q}_+$ and $\hat{q}_-$ respectively, can tune energy gaps between $\ket{A},\ket{B}$ and $\ket{TT}$ states to mix electronic degrees of freedom. This vital mechanistic significance of in-phase vibrational motions in SEF has been unrecognized in literature, except the adiabatic vibronic coupling density analysis of Nakano et al. \cite{Nakano2015} who remark on the importance of high-frequency in-phase vibrational motions. For example, recent two-state models\cite{Rao2016,Miller2020} of SEF will miss this central role of in-phase vibrational motions. The mechanistic significance of the adiabatic inseparability of both $\hat{H}(\hat{q}_-)$ and $\hat{H}(\hat{q}_+)$ from the electronic Hamiltonian $\hat{H}_{elec}$ in Equation~\ref{eq2} is illustrated below in Section \ref{qplusminuscouple}.\\

%Unlike two-state descriptions, the SEF Hamiltonian in Equation~\ref{eq2} considers all five possible electronic states in the one-quantum manifold arising in a dimer. This leads to five diabatic excitons upon diagonalization of  $\hat{H}_{elec}$. 

%The key point in Figure~\ref{fig:fig1}a is that vibronic mixing \textit{promoted} by correlated vibrational motions in the SEF Hamiltonian is an additional effect compared to an excitonic dimer, where correlated vibrations do not mix electronic degrees of freedom and are mere spectators. 
%as well as the vibrational dimensionality of the problem by making one-particle approximations.  \textit{The above physical picture also suggests that any possibility of an intermolecular conical intersection in SEF, as invoked in recent literature\cite{Miller2020}, has to allow for both correlated and anti-correlated vibrational modes}.  

\subsection{In-phase and Anti-phase Motions Create Coupled Vibronic Resonances}\label{qplusminuscouple}
Prominent vibrational quantum beats in 2DES experiments on acenes, both monomers\cite{Tan2021} and thin films\cite{Rao2016,Duan2020}, which match FC vibrations in the Raman spectra are well-known. With dimensionless FC displacements close to unity\cite{Hestand2018}, large FC displacements in acenes are strikingly different from those in photosynthetic pigments. Larger widths of vibronic resonances, proportional to $\sim \omega d \sin(2\theta_d)$ for an excitonic dimer\cite{Tiwari2018,Sahu2020}, may therefore be naturally expected in case of acene-based systems. $\theta_d$ is the electronic mixing angle between electronic states $\ket{A},\ket{B}$ which is also expected to be larger in SEF. Below we illustrate how these unique features influence vibronic couplings in SEF.
%Below we show that presence of multiple excitons, vibronic mixing along both $\hat{q}_+$ and $\hat{q}_-$ and large expected widths of vibronic resonance ensures that exact matches between excitonic energy gaps and vibrational frequencies are not necessary to promote vibronially enhanced SEF, and that multiple vibronic resonances, individually along $\hat{q}_+$ and $\hat{q}_-$, are likely. We illustrate these previously unrecognized effects in SEF by first considering a high frequency intramolecular vibrational mode in the SEF Hamiltonian in Equation~\ref{eq1}.  \\

We consider a  high-frequency intramolecular FC mode of frequency $\omega_H = $1360 cm$^{-1}$ as part of the system Hamiltonian, where numerical diagonalization of the Hamiltonian ensures that its non-adiabatic mixing with electronic degrees of freedom is treated exactly\cite{Tiwari2013,Peters2017} without making Born-Oppenheimer approximation. This mode corresponds to the symmetric stretching mode of the pentacene ring also included in prior studies\cite{Tempelaar2017_1,Hestand2015}. The remaining vibrational bath is modeled as temperature-dependent underdamped Brownian oscillators which introduce decoherence\cite{Tiwari2017} between the ground and excited state electronic energy gap, and consequently homogeneously broadened lineshapes. Parameters for Brownian oscillator lineshapes are derived from fits of experimental linear absorption spectrum of pentacene monomers in ref.\cite{Zhao2021}. The total Huang-Rhys (HR) factor in the high-frequency modes on the locally excited states is unity to be consistent with prior models\cite{Hestand2018,Tempelaar2017_1}. Cationic and anionic stabilization energies for pentacene are approximately equal \cite{Bredas2007}, as are the respective vibrational frequencies on cationic and anionic transient states\cite{Frontiera2018}. Parameters for the electronic Hamiltonian have been taken from ref.\cite{Martinez2020} which serve as a good approximation to a higher-level calculation reported\cite{Ananth2014} earlier by Ananth et al. The five electronic states, without any vibrations, undergo electronic mixing to produce five excitons which are denoted as $\epsilon$, $\alpha$, $\beta$, $\gamma$, and $\delta$ according to increasing excitonic energy. The basis state character of these excitons is listed in \textcolor{black}{Table S5}. The two highest energy excitons ($\gamma, \delta$) in the dimer are CT-dominant while the lowest exciton $\epsilon$ is $TT$-dominant with $\sim$7\% total bright character from locally excited states. The lineshapes are ensemble averaged over an energetic disorder of $\sigma = $ 75 cm$^{-1}$ between excited electronic states, estimated by Rao et al. from fits of 2DES simulations to experimental data. The resulting ensemble averaging only affects experimentally observed dephasing of excited state coherences\cite{JonasARPC2018} (shown later in Sections \ref{enhance} and \ref{2d}). Inhomogeneity in the 0--1 optical energy gap only broadens the lineshapes without affecting decoherence or dephasing rates. This is not included in order to highlight the rich underlying spectroscopic features. Calculations including the 0--1 inhomogeneity are shown in \textcolor{black}{Sections S6}. Further details of model parameters are discussed in \textcolor{black}{Section S3}. Convergence checks are described in \textcolor{black}{Section S5}.

%of $\sim$300 cm$^{-1}$
%Petelenz et al. have estimated $\sim$500 cm$^{-1}$ of correlated inhomogeneity in the 0--1 optical energy gap in pentacene thin films. Such an energetic disorder does not affect ensemble dephasing and therefore not introduced in the simulations}.
 %Prior works\cite{Martinez2020,Berkelbach3,Tempelaar2018_3} have reported that \% CT character in the lowest three excitons tends to increase in an extended crystal with greater number of CT manifolds.
%Thus in our model, we can calculate the ratio of locally excited state displacement and charge transfer state (CT) displacement. This ratio has been used throughout to calculate CT displacements for different modes of pentacene, whose HR factors were obtained from experimental resonance Raman spectrum\cite{Rao2016}.
%\vt{However, this CT mediated mixing between $TT$ and $A,B$ states is by itself insufficient\cite{Tempelaar2017_1} to explain the efficiency and timescales of the SEF process.} 

The consequences of broad vibronic resonance widths expected in acene-based systems are understood by analyzing the oscillator strengths beneath the simulated absorption lineshapes in Figure \ref{fig:fig1}b. The line strengths in the dimer are sufficiently complicated even with one intramolecular vibrational mode. However, a pattern of three unequal intensities, marked as faint gray band, arises due to vibronic mixing  between bright $\alpha$ exciton and a quantum of vibrational excitation on $\epsilon$, denoted as $(10)_{\epsilon\alpha}$ near-resonant manifold. Peaks in the $\sim$15200-16550 cm$^{-1}$ region arise from multiple vibronic mixing channels. Exciton $\beta$ is near-resonant with $\alpha$ excitons with one quantum of vibrational excitation. This latter manifold is also near-resonant with the second vibrational excitation on $\epsilon$, together leading to a complex set of line strengths. This manifold is denoted as $(210)_{\epsilon\alpha\beta}$. Similar patterns have been semi-analytically described\cite{Tiwari2018,Sahu2020} in case of photosynthetic dimers. 

To confirm the above intensity borrowing effects, we analyze the \textit{widths} of these near-resonances. Jonas and co-workers have defined\cite{Tiwari2018} a width of vibronic resonance based on the range of frequencies over which intensity borrowing ratio changes from 1:2 to 2:1. For a photosynthetic dimer, expected resonance widths are only $\sim$30 cm$^{-1}$. In comparison when intensities of peaks labelled `1' and `2' in $(10)_{\epsilon\alpha}$ are plotted as a function of $\omega_H$, a $\sim$10x larger width of 330 cm$^{-1}$ is seen in Figure \ref{fig:fig1}c. When peaks in $(210)_{\epsilon\alpha\beta}$ manifold are analyzed while the vibrational frequency is scanned, a similar intensity borrowing pattern between `3' and `4' is seen with even broader widths. This confirms the expectation that non-adiabatic vibronic mixing in case of acenes is highly robust to resonance mismatches, and therefore becomes \textit{unavoidable} over a broad range of $\omega_H$ vibrational frequencies over which the optically bright excitons intermix, as well as impart bright character to vibrationally excited $TT$-dominant $\epsilon$ exciton. The physical picture that emerges from the above discussion is summarized in Figure~\ref{fig:fig1}d. Broad resonance widths are vital for enabling bright excitons ($\alpha-\beta$) intermixing through anti-phase motions along $\hat{q}_-$, as well as their mixing with $\epsilon$ through in-phase motions along $\hat{q}_+$, both caused by a \textit{single} intramolecular vibration on each molecule, thereby leading to coupled vibronic near-resonances as shown in Figure~\ref{fig:fig1}d. \textit{Importantly, two-state models with reduced vibrational dimensionality\cite{Rao2016,Duan2020} will miss out on these previously unrecognized effects that are unique to the SEF Hamiltonian. The mechanistic significance of robust, coupled vibronic resonances and the complementary role of low-frequency vibrations in this picture is elucidated in Section \ref{peierls}.}  \\

%The physical picture that emerges from the above discussion is summarized in Figure~\ref{fig:fig1}d.  The $(10)_{\epsilon\alpha}$ vibronic mixing primarily driven by $\hat{q}_+$ and $(10)_{\alpha\beta}$ vibronic mixing driven by $\hat{q}_-$ together lead to a coupled resonance $(210)_{\epsilon\alpha\beta}$ that not only mixes the bright excitons together but also leads to increased mixing of $TT$ and locally excited states $A,B$. This enhanced mixing of $TT$ will be recalled again in Section \ref{enhance} in the context of enhanced vibronic delocalization and population transfer. While $(10)_{\epsilon\alpha}$ kind of resonance is well-known in case of photosynthetic excitons and also discussed by Tempelaar and Reichman\cite{Tempelaar2018_3} to explain enhanced SEF rates, \textit{the vital role of correlated vibrational motions, coupled resonances arising from $\hat{q}_+$ and $\hat{q}_-$ motions, and broad resonance widths highlighted here are new and unique to the SEF Hamiltonian.} 

%\ab{@AB-q+ couples alpha and epsilon,q- couples beta-alpha and beta-epsilon domains(from 3 state matrices.)}

\subsection{Peierls Coupling Complements Vibronic Resonance}\label{peierls}
 Micro-Raman spectroscopy\cite{Venuti2002,Venuti2004} of acene thin films has reported several prominent low-frequency vibrations of mixed inter/intramolecular character which cause\cite{Troisi2006,Nakano2015} non-perturbative modulations of orbital overlap transfer integrals to affect intermolecular Peierls coupling. Huo et al. have suggested\cite{Huo2017} that a distribution of electronic couplings created by such low-frequency vibrations prevents destructive interference between CT couplings to enhance SEF. On the contrary, Herbert et al. have suggested\cite{Herbert2017} only a minor role for such couplings in enhancing SEF, with dominant role played by linear vibronic couplings arising from high frequency C--C stretching motions. Similarly, Tempelaar et al. have reported\cite{Tempelaar2018_3} that Peierls coupling `breaks' vibronic resonance condition to diminish SEF rates. In addition to the above inconsistent understanding of the role of Peierls coupling in SEF, models so far have suggested exclusive roles for Peierls coupling and resonant vibronic couplings. Here we will illustrate the complex interplay of these couplings to introduce new insights in this picture. \\

Peierls coupling is introduced through a low frequency vibration $\omega_{L} = 265$ cm$^{-1}$, prominent in 2DES\cite{Rao2016} and resonance Raman\cite{Venuti2002,Venuti2004} experiments. The corresponding vibrational coordinates are $\hat{q}_{A(B)_L}$ with FC displacements $d_L$ on locally excited states $\ket{A},\ket{B}$, $d_L^{TT}$ on $\ket{TT}$ state and $d_L^{CT}$ on the $\ket{CT}$ states.  Modulation of overlap integrals due to the low-frequency motions is additionally incorporated through Peierls coupling Hamiltonian\cite{Bredas2007} $\hat{H}_P = \sum_{J \neq K}\kappa(\hat{q}_{A_L} + \hat{q}_{B_L}) \ket{J}\bra{K}$. Indices $J$ and $K$ run over all the electronic states similar to $\hat{H}_{elec}$, and $\kappa$ denotes the coupling constant. Based on the calculations in ref.\cite{Duan2020}, $\kappa$ is estimated to be $\kappa = $150 cm$^{-1}$. As before, all the parameters are directly motivated from experiments as described in \textcolor{black}{Section S3 and listed in Table S4}. Now both $\omega_H$ and $\omega_L$ vibrations are treated as quantum oscillators in the system Hamiltonian such that each diabatic exciton potential becomes 4-dimensional with diabatic basis states denoted as $\ket{X}\ket{\nu^X_{A_H}}\ket{\nu^X_{B_H}}\ket{\nu^X_{A_L}}\ket{\nu^X_{B_L}}$. $X$ runs over all five excitons, with $\nu$ quanta of vibrational excitation. Note that, as before, ground state vibrations of the unexcited molecule are still allowed to be a part of an overall excited electronic state, that is, the basis set is numerically exact without any 1PA. It is crucial to also note that Peierls distortion has been treated as a quantum oscillator in the system Hamiltonian instead of electronically off-diagonal disorder under the `frozen mode' approximation\cite{Castillo2015,Tempelaar2018_3}. Although this makes the four-wavemixing simulations significantly more expensive, we will show that quantum oscillator treatment is \textit{necessary} in order to describe the complementary nature of Peierls coupling. \textit{To best of our knowledge, such effects have not been reported earlier because of either exclusive treatment of both couplings, or approximating low-frequency Peierls distortion as a 'frozen' mode.} \\

%$\omega_L$ is a representative mode for mixed inter- and intramolecular low-frequency vibrations reported in resonance Raman\cite{Venuti2002,Venuti2004} and 2DES spectra\cite{Rao2016} of pentacene thin films. The FC displacements are deduced from relative intensities of $\omega_H$ and $\omega_L$ modes in the experimental resonance Raman spectrum\cite{Rao2016}.

%The FC displacements on the locally excited states, $d_L$ are deduced from intensities of the corresponding mode in resonance Raman spectrum\cite{Rao2016}. FC displacements on all other electronic basis states are estimated in the same way as those for $\omega_H$ in Section \ref{qplusminuscouple}.

\begin{figure*}[ht]
	\centering
	\includegraphics[width=5 in] {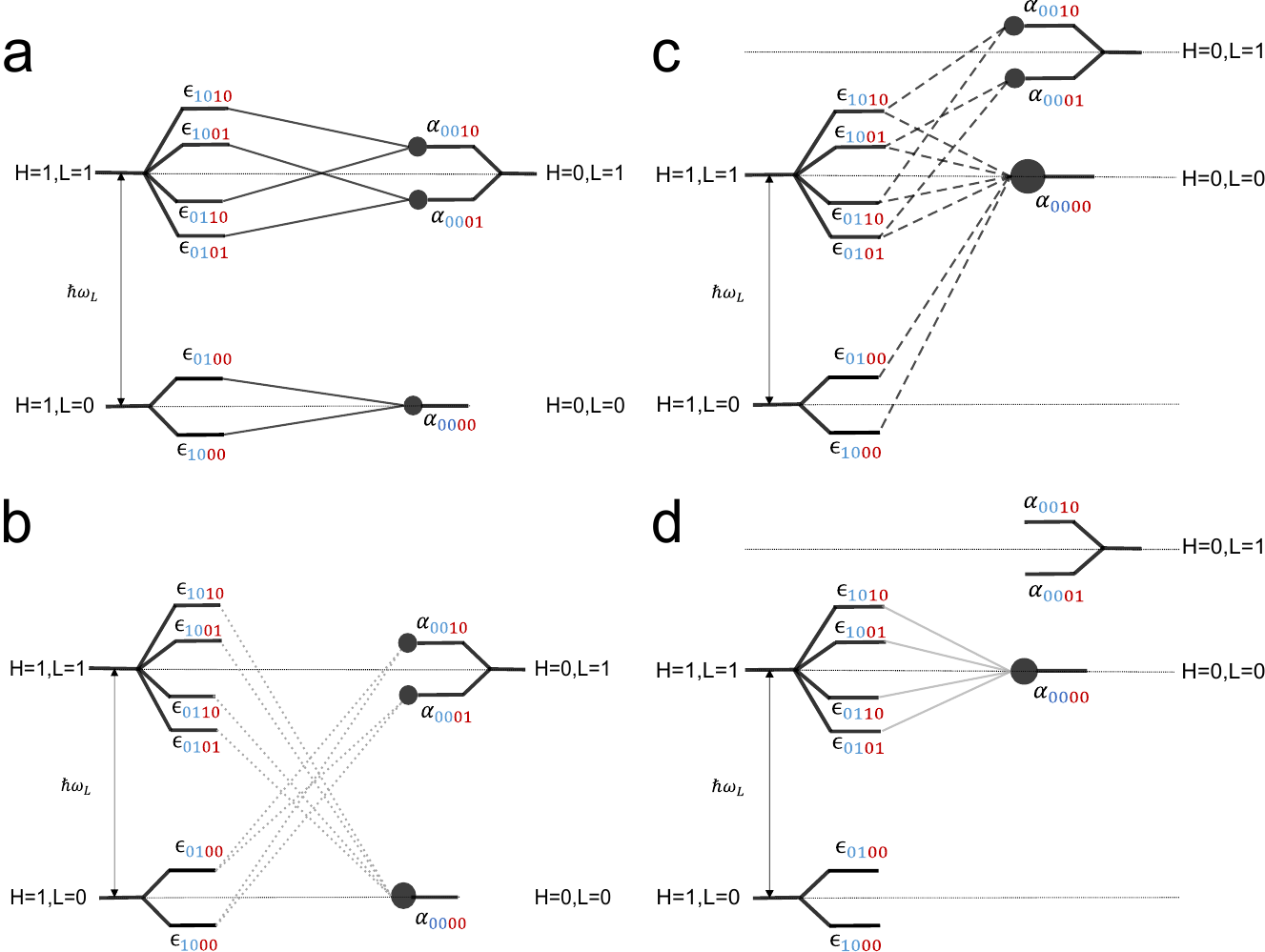}
	\caption{\footnotesize Peierls coupling complements resonant vibronic mixing. A schematic showing the dominant linear vibronic (panels a,c) and Peierls (panels b,d) couplings between diabatic excitonic energy levels in the $(10)_{\epsilon\alpha}$ near-resonant manifold. The basis states are represented as $X_{\nu_{A_H}\nu_{B_H}\nu_{A_L}\nu_{B_L}}$, where $X$ denotes diabatic exciton and $\nu$ denotes quanta in the respective vibrational basis states. Dominant coupling matrix elements, those without any energetic mismatch and 0-1 FC overlaps, are denoted by solid lines, while those with both or any one unfavorable factor are denoted by dotted and dashed lines, respectively. $H(L)$ excitation denotes excitation of $\omega_{H(L)}$ vibration on any one of the molecules $A, B$. The resulting set of degenerate levels are offsetted for clarity. Size of the solid black dot on $\alpha$ exciton denotes the number of coupling channels associated with it. (A) Linear vibronic couplings for exact $(10)_{\epsilon\alpha}$ resonance. (b) Peierls couplings for exact $(10)_{\epsilon\alpha}$ resonance. (c) Linear vibronic couplings for $(10)_{\epsilon\alpha}$ near-resonance such that lack of resonance is compensated by excitations along the low-frequency vibration, that is, $L=1$. (d) Peierls couplings for the near-resonant case. Minor interactions mediated by Peierl's coupling such as between $\epsilon_{1010}$ and $\alpha_{0010,0001}$ are not shown as they are associated with weak FC overlap factors. Note that the linear vibronic couplings arising from the FC active low-frequency vibration are not shown. \textcolor{black}{See Section S1 for more details.}}
	\label{fig:fig2}
\end{figure*}
\FloatBarrier
\normalsize

%%%% Comment about possibility of secondary interactions %%%%%
%\ab{AB, I think in panel B couplings such as $\epsilon_{1010}$ and $\alpha_{0010,0001}$ are also possible. And these will also have two unfavorable factors (dotted), the 0-1 overlap along $\hat{q}_{A_H}$, and 1-$\hat{q}_{A_H}$-1 or 0-1 factor along $\hat{q}_{B_L}$. To decide whether to show these, compare $\braket{1^{\epsilon}_{A_L}|\hat{q}_{A_L}|1^{\alpha}_{A_L}}$ and $\braket{1^{\epsilon}_{A_L}|\hat{q}_{A_L}|0^{\alpha}_{A_L}}$ carefully. Along similar lines, in panel D, couplings between $\epsilon_{1010}$ and $\alpha_{0010,0001}$ are possible, although the most dominant channel is already shown in the figure.

The interplay of linear vibronic and Peierls coupling can be analyzed by considering the two scenarios shown in Figure~\ref{fig:fig2}, where the exciton basis states are represented as \small $X_{\nu_{A_H} \nu_{B_H} \nu_{A_L} \nu_{B_L}}$ \normalsize for brevity. The left column, panels A and B, considers exact resonance due to $\omega_H$, while the right column, panels C and D, considers near-resonant $\omega_H$ where the lack of resonance is compensated by a quantum of excitation along the low frequency vibration $\omega_L$. In each column, the top and bottom panels show the dominant linear vibronic and Peierls coupling channels, respectively. Considering Figure~\ref{fig:fig2}a, one of the dominant linear vibronic coupling matrix element in the $(10)_{\epsilon\alpha}$ manifold, denoted as $H=1, L=0$, will be proportional to --\\
\footnotesize $J_{LV} \propto \braket{1^{\epsilon}_{A_H}|\hat{q}_{A_H}|0^{\alpha}_{A_H}} \braket{0^{\epsilon}_{B_H}|0^{\alpha}_{B_H}}\braket{0^{\epsilon}_{A_L}|0^{\alpha}_{A_L}}\braket{0^{\epsilon}_{B_L}|0^{\alpha}_{B_L}}$. \normalsize \\ A similar matrix element can be written along $\hat{q}_{B_H}$. Non-dominant coupling elements will also arise due to linear vibronic coupling along the low-frequency FC modes $\hat{q}_{A(B)_L}$, and not shown in the figure. Interestingly, panel A shows that a quantum of excitation along the low-frequency mode, for example $\hat{q}_{A_L}$, leads to identical resonance conditions in the $H=1,L=1$ manifold with twice as many coupling channels, although with an accompanying vibrational overlap factor $\braket{1^{\epsilon}_{A_L}|1^{\alpha}_{A_L}}$. These factors are approximately the same as those arising from $\braket{0|0}$ overlaps owing to large FC displacement in acenes. \textcolor{black}{See Section S1 for calculations of these FC overlaps}. Thus, presence of low-frequency FC active vibrations by themselves enhance resonant vibronic mixing by increasing the number of mixing channels. \\

Figure~\ref{fig:fig2}b shows that when Peierls coupling is introduced in the above resonant picture, additional coupling channels open up in the $(10)_{\epsilon\alpha}$ manifold. For example, one of the dominant Peierls coupling matrix element $J_P$ between $H=1,L=1$ and $H=0,L=0$ along $\hat{q_{A_L}}$ will be proportional to --\\
\footnotesize $J_P \propto \braket{1^{\epsilon}_{A_H}|0^{\alpha}_{A_H}} \braket{0^{\epsilon}_{B_H}|0^{\alpha}_{B_H}}\braket{1^{\epsilon}_{A_L}|\kappa \hat{q}_{A_L}|0^{\alpha}_{A_L}}\braket{0^{\epsilon}_{B_L}|0^{\alpha}_{B_L}}$. \\
\normalsize Note that energetic mismatch between these participating levels will suppress perturbative mixing between these states caused by Peierls coupling. The two unfavorable factors, weaker mixing and 0-1 FC overlaps, are denoted by dotted lines in the panel. However, increased number of mixing channels and substantial FC overlaps due to large intramolecular FC displacements in acenes (unlike in the case of photosynthetic excitons) together ensure that their net effect is consequential (Section \ref{enhance}). \\

Figures~\ref{fig:fig2}c,d show how the resonant picture modifies when there is energetic mismatch in the $(10)_{\epsilon\alpha}$ manifold. The dominant coupling matrix elements are marked on the figure. Compared to panel A, the near-resonant vibronic mixing in panel C is weakened (shown in dashed lines) due to energetic mismatch. However, a low frequency vibrational excitation makes up for the mismatch and opens up additional mixing channels between \small $\alpha_{0000}$ \normalsize and four possible degenerate states with $H=1,L=1$ excitation on $\epsilon$. These channels are denoted as dashed line because a 0-1 FC overlap reduces the mixing. With the same reasoning, a comparison of energetic alignment between panels B and D shows that Peierls couplings of panel B become \textit{stronger} in panel D because the earlier energetic mismatch is now compensated for by low-frequency vibrational excitations. It is important to note that without a vibronic near-resonance, low-frequency vibrational excitations will not by themselves be sufficient to compensate for large energetic mismatch between exciton energies of (TT)$^1$ and S$_1$ manifolds. A subtle point to also note is that in case of only minor energetic mismatches \textit{introduced} by low-frequency vibrational excitations, increase in the number of mixing channels especially in the presence of Peierls coupling, can more than compensate for it, biasing the overall effect towards enhancement of vibronic mixing.  In summary, \textit{Peierls coupling through low-frequency vibrations increases the number of vibronic mixing channels and is expected to complement resonant vibronic mixing by making it more robust to energetic mismatches}. \\

Based on above considerations, the width of vibronic resonance (Figure~\ref{fig:fig1}b), exciton delocalization and population transfer rates from the locally excited states $\ket{A},\ket{B}$ to the acceptor state $\ket{TT}$, are all expected to be enhanced. These effects are confirmed through numerical calculations in Section \ref{enhance}, and are contrary to earlier treatments\cite{Tempelaar2018_3, Herbert2017, Duan2020} of Peierls coupling where, either an explicit quantum oscillator treatment of low-frequency vibrations was not possible, or Peierls coupling was treated exclusively of linear vibronic coupling.

%\ab{What does Nakano say about Peierls coupling, does he have to be cited in the preceeding line ?
%Nakano has done a vibronic coupling density analysis by expanding the hamiltonian in terms of Q, where off-diagonal elements are Peierl's coupling and diagonal ones are LVC. So, to my understanding they have taken both the couplings into account and given a list of couplings for in phase and out-of-phase vibs for a range of energy 50-1700 cm$^{-1}$. Although their system is teracene I think we should cite them here.} . 

 \subsection{Enhanced Vibronic Mixing, Resonance Width and Exciton Delocalization}\label{enhance}
 Intensity borrowing effects in oscillator strengths, such as those in Figure~\ref{fig:fig1}b can be directly analyzed to understand delocalization. However oscillator strengths even for the lowest energy excitons are fairly complicated when both $\omega_{H}$ and $\omega_L$ intramolecular vibrations are introduced in the Hamiltonian (\textcolor{black}{see Figure 3a}). Instead, we use inverse participation ratio (IPR) as an alternative metric to analyze these effects. IPR is a measure of excitonic delocalization which has been previously extended\cite{Womick2011} by Moran et al. to quantify delocalization in vibronic excitons. A lower IPR implies more exciton delocalization with perfect delocalization corresponding to an IPR of $1/N$, for $N$ electronic states. \\
 
%$\hat{H}_{elec}$ in the SEF Hamiltonian suggests only weak mixing between $TT$ and locally excited $A,B$ states. Note that perfect delocalization in this Hamiltonian corresponds to an \vt{IPR of 0.2 which is quite close to the IPR for purely electronic exciton $\alpha$ which lies at 0.298.} Similarly, the IPR for the $TT$-dominant exciton $\epsilon$ is 0.67 with \vt{at least three of the brightest oscillator strengths localized equal} or more than $\epsilon$.
 
\begin{figure*}[h!]
 	\centering
 	\includegraphics[width=3 in]{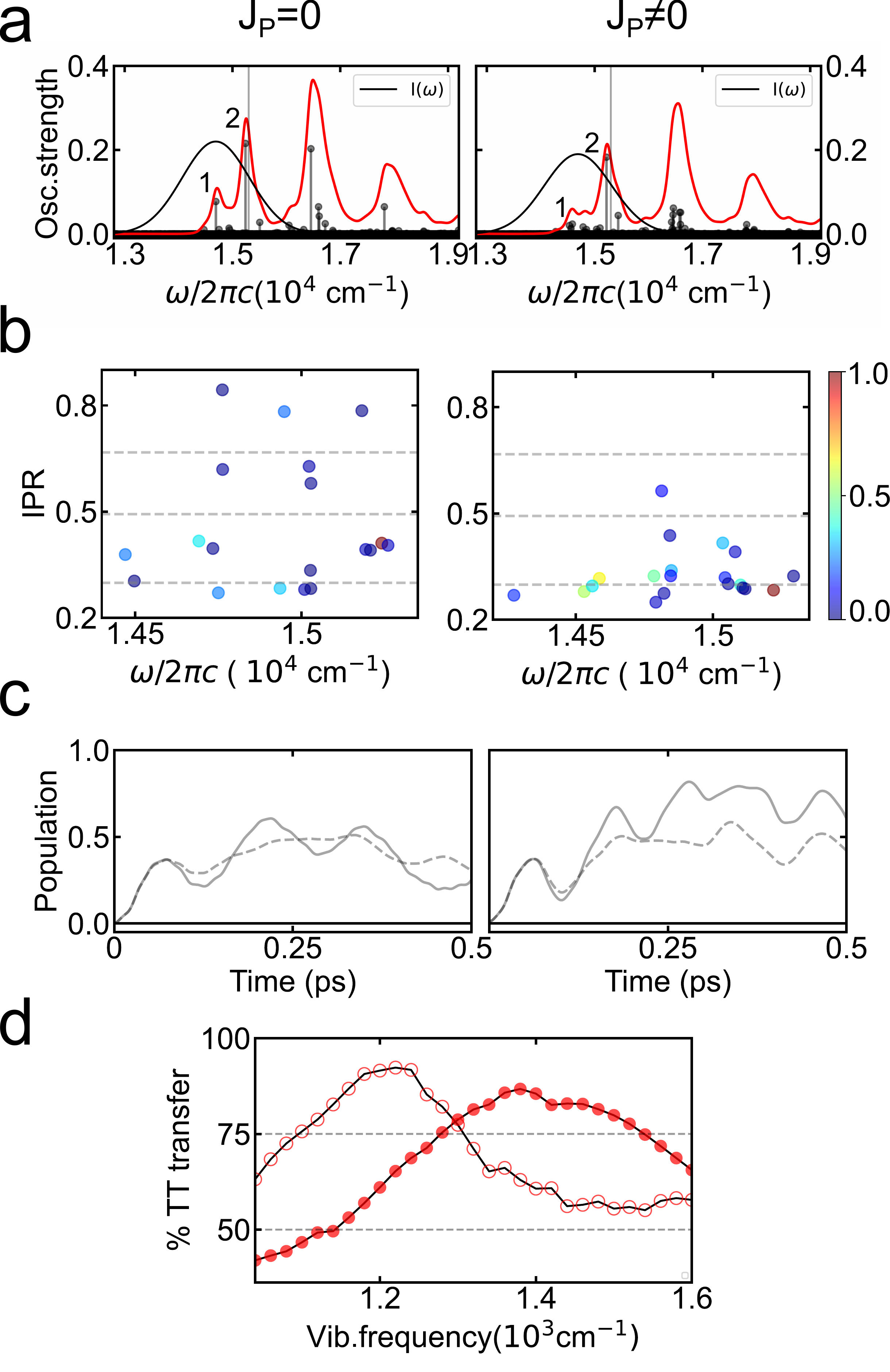}
 	\caption{\footnotesize (\textbf{a}) Exact calculation of absorption intensities of the pentacene dimer at 300 K	q without (left) and with (right) Peierls coupling. The line strengths are overlaid with Brownian oscillator lineshapes. Each molecule has two Franck-Condon active vibrations, 1360 cm$^{-1}$ and 265 cm$^{-1}$, included explicitly in the system Hamiltonian. The intensity spectrum of the laser pulse used in 2D simulation is overlaid as black solid line. The faint vertical line shows the upper limit of the square pulse used in wavepacket dynamics simulations. Transitions marked by numbers are also overlaid in the CMs and 2D spectra in Section \ref{2d}. Details of the calculation are mentioned in \textcolor{black}{Section S3} .(\textbf{b}) Inverse Participation Ratio (IPR) of the 20 brightest excitons in the energetic range up to $\sim$15300 cm$^{-1}$ covering mixed $\epsilon$, $\alpha$ and $\beta$ vibronic eigenvectors. The oscillator strengths are chosen from the simulation of linear absorption strengths in \textcolor{black}{ Figure \ref{fig:fig3}a}, represented as color map, and plotted as a function of energy and IPR. For the sake of clarity, oscillator strengths of the brightest two eigenvectors in the left panel and the brightest one in the right panel have been scaled down by a factor of 5. The IPRs expected from the purely electronic Hamiltonian without vibronic mixing are denoted as dashed gray lines, and correspond to $\epsilon$, $\beta$ and $\alpha$ excitons with decreasing IPRs, respectively. Left (right) panel corresponds to the case of without (with) Peierls coupling introduced through $\omega_L$. (\textbf{c}) Excited state wavepacket dynamics following an impulsive excitation of a superposition of vibronic eigenvectors resulting from the exact non-adiabatic SEF Hamiltonian with $\omega_H$ and $\omega_L$ intramolecular vibrations. Left (right) panels correspond to $\ket{TT}$ acceptor population post photoexcitation without (with) Peierls coupling. Dashed curves correspond to ensemble averaged dynamics introduced by an anti-correlated energetic disorder\cite{Rao2016}, $\sigma =$ 75 cm$^{-1}$ between $\ket{TT}$ and locally-excited $\ket{A},\ket{B}$ states The impulsive excitation corresponds to a spectrally flat laser pulse polarized along $\alpha$ exciton, centered at 14150 cm$^{-1}$ and covering eigenvectors in the $\epsilon, \alpha, \beta$ manifold up to 15300 cm$^{-1}$ energy as shown in \textcolor{black}{Figure 3a}. Full details of the calculation are presented in \textcolor{black}{Section S3}. (\textbf{d}) \% population transfer to the $\ket{TT}$ plotted as a function of $\omega_H$ vibrational frequency corresponding to non-ensemble averaged dynamics in panel b. Hollow circles correspond to no Peierls coupling (panel B left) and solid circles correspond to the calculation with Peierls coupling (panel B right)..
 	}
 	\label{fig:fig3}
 \end{figure*}

In Figure~\ref{fig:fig3}a shows the exact calculation of absorption intensities of a pentacene dimer at 300 K without(left panel) and with(right panel) Peierls coupling. Each molecule has two FC active vibrations -- 1360 cm$^{-1}$ and 265 cm$^{-1}$. It is evident that with the inclusion of Peierls coupling, oscillator strengths are further distributed among more number of excitons indicating larger delocalization due to increase in inter-exciton mixing channels. The black Gaussian curve covering the low-energy excitons shows the intensity of the laser pulse used for 2D spectroscopy simulations (shown later in Section \ref{2d}). The absorption sticks are overlaid with the lineshapes obtained from underdamped Brownian quantum oscillators\cite{Mukamel1990}. Further details of the calculation are provided in \textcolor{black}{Section S3}. \\

 Enhanced delocalization due to Peierls coupling is further confirmed in Figure~\ref{fig:fig3}b by comparing IPR of brightest 20 vibronic eigenvectors in a range of $\sim$2300 cm$^{-1}$ within the lower energy manifold. IPR is compared for the case of with versus without Peierls coupling ($\kappa = 0$). The oscillator strengths are chosen from the simulation of linear absorption strengths in \textcolor{black}{Figure~\ref{fig:fig3}a}. The dashed horizontal lines in Figure~\ref{fig:fig3}b are the IPRs for the lowest three excitons ($\epsilon$, $\alpha$, and $\beta$) of the purely electronic Hamiltonian $\hat{H}_{elec}$, and shown for reference. From a comparison of left and right panels in Figure \ref{fig:fig3}b, it is evident that exciton delocalization is enhanced in the presence of Peierls coupling such that optically bright excitons are also more delocalized on average. Increased delocalization of optically bright excitons implies that optical excitation will directly create vibronic wavepackets with substantial character from $\ket{TT}$ states. \\

%Following the 2DES spectroscopic simulations of Rao et al. \cite{Rao2016} to fit experimental 2D spectra and coherence maps (CMs), we introduce (\ab{ $\sigma$=75cm$^{-1}$})anti-correlated energetic disorder between the site energy of $\ket{TT}$ and $\ket{A},\ket{B}$ states. s, \sout{Amitav double check once, seems long}} even if no explicit zero-quantum decoherence mechanism is introduced. Such a decoherence mechanism is not amenable\cite{JonasARPC2018} to direct experimental probe due to ensemble dephasing affecting the measured dephasing rates.
 
 In the Fermi's Golden rule rate limit, increased vibronic mixing channels (Figure \ref{fig:fig2}) imply faster SEF rates with couplings $J_c$ through individual mixing channels adding\cite{Tiwari2017,Patra2022} as $J^{TOT} = \sqrt{\sum_{c} J_c^2}$. Increased delocalization of vibronic excitons in the presence of Peierls coupling from low-frequency vibrations also suggests enhanced population transfer to $\ket{TT}$ when a \textit{coherent} superposition of bright excitons is impulsively excited. This is confirmed in Figure~\ref{fig:fig3}c which compares coherent population transfer to $\ket{TT}$ expected from the system Hamiltonian following an impulsive photoexcitation polarized along the $\alpha$ exciton. \textcolor{black}{The calculation compares the case of with versus without Peierls coupling with full details in Section S2}. Including energetic disorder\cite{Rao2016} in site energies causes disorder in excitonic energy gaps leading to ensemble dephasing\cite{JonasARPC2018} of electronic and vibronic quantum beats, with \vt{1/e} timescale\cite{JonasARPC2018} of $\sim$100 fs. In case of no Peierls coupling (left panel), $\sim$66\% of the initial excitation of optically bright excitons is transferred to the optically dark $\ket{TT}$ states within $\sim$300 fs.  This is true even with ensemble averaging which suppresses purely electronic and vibronic quantum beats. The slower beats at 54 cm$^{-1}$ correspond to vibronic splittings similar to those in Figure \ref{fig:fig1}c, while the faster beats at 265 cm$^{-1}$ correspond to $\omega_L$ intramolecular vibration. Corresponding Fourier transforms are shown in \textcolor{black}{Figure S3}. The high frequency vibrational beat i.e. 1360 cm$^{-1}$ cannot be observed because $\omega_H$ excitations lie outside the laser bandwidth. In contrast, when Peierls coupling is introduced through the low-frequency vibration, the transfer is enhanced to $\sim$86\%, in line with the theoretical expectation from Figure \ref{fig:fig2}. Interestingly, ensemble dephasing strongly \textit{masks} this enhancement with apparent transfer approximately similar to that seen without Peierls coupling in the low-frequency mode, even though on an individual system level Peierls coupling and vibronic near-resonance together enhance population transfer. In contrast to vibronically enhanced transfer, a purely electronic picture, with parameters derived from ab-initio models\cite{Martinez2020}, predicts only 32\% of transfer when a coherent superposition of $\epsilon$, $\alpha$ and $\beta$ excitons is excited (\textcolor{black}{Figure S2}). When Peierls coupling is introduced without a high-frequency near-resonant vibration, population transfer no greater than that possible with purely electronic coupling is seen (\textcolor{black}{Figure S2}), underscoring the necessity of a vibronic near-resonance for Peierls coupling to be effective. Interestingly, the complementary effect of low-frequency modulations of intermolecular orbital overlaps in enhancing $\alpha-\beta$ mixing, also implies enhanced indirect $\epsilon-\beta$ mixing. This is evident from \textcolor{black}{Figure S4 and Table S2} with $\sim$2x increase in population transfer to $\ket{TT}$ upon selective excitation of the higher energy optically bright $\beta$ exciton.
 
 The synergistic effect of low and high frequency vibrations on vibronic near-resonance is further confirmed by analyzing the effect of Peierls coupling on the width of vibronic resonance. Presence of low-frequency vibrational mode significantly complicates the interpretation of oscillator strengths that was possible with only high-frequency mode in the Hamiltonian (compare Figure~\ref{fig:fig1}b with oscillator strengths plotted in \textcolor{black}{Figure \ref{fig:fig3}a}). Therefore, we use vibronically enhanced population transfer as a proxy for near-resonant intensity borrowing. Figure~\ref{fig:fig3}c plots the \% transfer as a function of $\omega_H$ which, as in Figure~\ref{fig:fig1}b, is scanned while keeping $\hat{H}_{elec}$ unchanged. The width of near-resonant vibrations over which transfer to $\ket{TT}$ state remains above 75\% increases by $\sim$1.24x, from 210 cm$^{-1}$ to 260 cm$^{-1}$. This increased width essentially implies that low-frequency modulations in intermolecular orbital overlaps in acene-based SEF systems make vibronic resonances, the width of which are already quite broad due to large HR factors (Figure~\ref{fig:fig1}b), highly robust to mismatches between vibrational frequencies and exciton energy gaps. \textit{This contrast between selective role of narrow vibronic resonances in photosynthetic excitons\cite{Engel2021,Tiwari2013} versus non-selective and robust vibronic resonances expected in SEF is quite striking. }

\subsection{Distinguishing Vibronically Enhanced From Spectator Quantum Beats Using 2DES}\label{2d}
Our findings suggest that strong mixing between excitons through vibrational motions may be expected in SEF upon photoexcitation. Such mixing of excitons can manifest as intensity borrowing effects in linear and non-linear optical spectra. \vt{Furthermore, when an impulsive photoexcitation creates a superposition of such vibronic eigenvectors, an amplitude enhancement of quantum beats on the excited and ground electronic state is expected}\cite{Tiwari2013,Tiwari2017,Tiwari2018}. Enhanced beating amplitude has been experimentally reported\cite{JonasARPC2018} in the context of pump-probe and 2DES experiments on photosynthetic excitons. Such quantum beats can serve as a sensitive probe of the underlying mechanism of energy/charge transfer. For example, vibrational wavepackets corresponding to motions which drive internal conversion through a conical intersection\cite{Yarkony1996} do not survive on the excited state while orthogonal motions do. In a square symmetric molecule, Jonas and co-workers have shown that\cite{Jonas2008,Jonas2008b,KitneyHayes2014} quantum beat amplitude of promoter vibrations is anisotropic to parallel versus perpendicular relative optical polarization of pump and probe electric fields. Polarization anisotropy of quantum beat amplitudes can therefore serve to distinguish promoter versus spectator vibrations. 

In the current context of SEF, our analysis suggests that vibronic states formed by mixing of excitons polarized along different directions will carry highly mixed optical polarization signatures. Furthermore, mixed polarization signatures of vibronic eigenvectors will be caused by not just high-frequency vibrations (as they lead to vibronic near-resonances), but also due to low-frequency vibrations due to their vital complementary role in enhancing vibronic mixing between excitons. However, considering these predictions in the light that large HR factors in acenes will by themselves lead to strongly modulated vibrational quantum beats in 2DES or pump-probe spectra, such as for pentacene monomers in solution\cite{Tan2021}, discriminating vibronically enhanced quantum beats against those arising from spectator vibrational motions becomes imperative in order to make sound judgements regarding the mechanistic details of SEF. Taking motivation from vibrational polarization anisotropy analysis\cite{Jonas2008,Jonas2008b,KitneyHayes2014} of Jonas and co-workers, below we propose a novel polarization-based 2DES approach for unambiguous spectroscopic identification of promoter versus spectator vibrational motions in SEF, suggesting experiments which can readily test the above mechanistic predictions.

%Crucially, the picture in Figure~\ref{fig:fig2} also predicts that Peierls coupling alone, without a vibronic resonance, will be ineffective in promoting vibronic mixing. 
%In the context of SEF in acene-based systems, our model Hamiltonian in Equation~\ref{eq1}, which treats a $\omega_H$ and $\omega_L$ vibration explicitly in the system Hamiltonian, predicts that large HR factors along with Peierls coupling will lead to a broad range of non-selective vibronic mixing promoted by both $\omega_H$ and $\omega_L$ intramolecular vibrations. 

Pump-probe spectroscopy  (PP) plots evolution of detection frequency $\omega_t$ of a system caused by pump excitation as a function of waiting time $T$ between pump and probe pulses. Akin to pump-probe, 2DES is also a four-wavemixing experiment. However, compared to PP, 2DES involves a pair of pump pulses, the delay $\tau$ between which is scanned in an interferometrically stable fashion to result in an additional Fourier transformed absorption frequency axis $\omega_{\tau}$. The resulting data is plotted as a 2D contour map that correlates excitation and detection frequency of the system. 2D snapshots as a function of $T$ report on rich vibrational-electronic relaxation dynamics of a system. In case of pentacene thin films, prominent quantum beats with a few picosecond dephasing timescale along $T$ are reported\cite{Rao2016,Duan2020,Musser2015} in 2DES and PP studies, and contribute as positive ground state bleach (GSB) and negative excited state absorption (ESA) signals. The excited state emission (ESE) signal is reported\cite{Friend2011} to exhibit a sub-100 fs decay due to rapid internal conversion of photoexcited singlet states into correlated triplets. Accordingly, following the 2DES simulations of Tempelaar and Reichman\cite{Tempelaar2017_2}, our 2DES calculations only consider the GSB and ESA signal contributions. In a broader context, it is well known that excited state wavepackets can be generated by impulsive internal conversion\cite{Joo2012}, can exhibit modified beating amplitudes\cite{Joo2012,Petelenz2019, Joo2020}, rapidly dephase upon internal conversion, or simply undergo coherence transfer\cite{Jean1995} to the product without promoting any vibronic mixing. Owing to these complications and the availability of clean GSB signals in case of SEF, we focus our analysis on the GSB signals arising from ground state vibrational quantum beats. Only rephasing 2DES signal contributions, where 2DES pulse sequence rephases the optical dephasing that occurred during the first time interval, are considered. Due to simultaneous availability of superior time resolution and excitation frequency information in 2DES, the polarization-based spectroscopic signatures of promoter modes that we propose will be based on 2DES. Equivalent PP signatures are possible, although may be ambiguous due to lack of excitation frequency information and no separation of rephasing and non-rephasing pathways. 

	\begin{figure*}[h!]
	\centering
	\includegraphics[width=4 in]{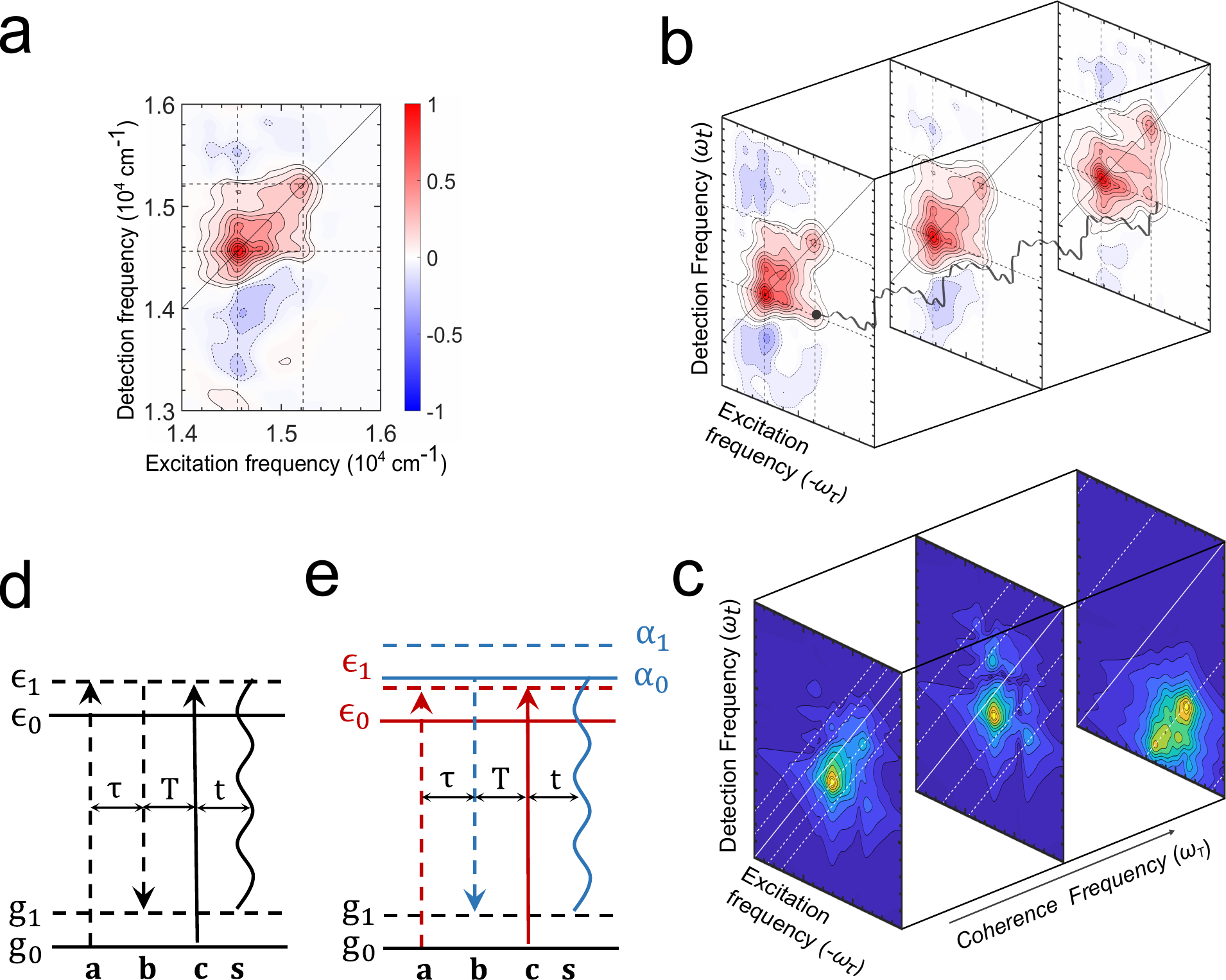}
	\caption{\footnotesize (\textbf{a})  Absorptive 2DES spectrum at $T =$ 100 fs simulated at 300 K, corresponding to the Hamiltonian in Equation~\ref{eq1} with $\omega_{H}$ and $\omega_{L}$ FC active intramolecular vibrations. Peierls coupling is introduced through the low-frequency vibration. The simulation parameters and details are described in Section S4. {Corresponding absorption lineshape with the laser spectrum overlaid is shown in Figure \ref{fig:fig3}a.} (\textbf{{b}}) Schematic description of 2DES quantum beat maps. Each $(\omega_{\tau},\omega_t)$ pixel on the 2DES map evolves with pump-probe waiting time $T$ due to incoherent population and coherence quantum beat pathways. The latter are isolated through a global fit across all pixels, and residuals Fourier transformed along $T$ to result in the coherence axis $\omega_T$. (\textbf{{c}})Coherence maps (CMs) stacked along $T$ axis. Each 2D CM is then a slice along a given $\omega_T$. (\textbf{{d}}) A given wavemixing diagram corresponding to ground state vibrational quantum beats arising in an isolated two-electronic level system with two vibrational quanta on each electronic state along a given intramolecular vibrational mode. Each such diagram corresponds to a given term in the third-order perturbative expansion of the time-dependent density matrix of the system and contributes at a given 2D location. The length of first and last interactions corresponds to excitation and detection frequency, respectively. The time interval between these light-matter intercations are shown($\tau$,T,t). The 2D location lies at $(\omega_{\tau}, \omega_t) = (\omega_{eg} + \omega_v, \omega_{eg})$, where $\omega_{eg}$ is the $e_0-g_0$ energy gap. Dashed and solid arrows indicate interactions from bra and ket side, respectively, while the last (fourth) interaction represents radiated electric field. (\textbf{{e}}) GSB quantum beat pathway arising from vibronic mixing. Color denotes different transition dipole character and directions, with red and blue corresponding to $\hat{\mu}_{g\epsilon}$ and $\hat{\mu}_{g\alpha}$ transition dipoles, respectively. The 2D location for this wavemixing diagram corresponds to the lower detection energy 2D cross peak between diagonal peaks corresponding to $\omega_{\epsilon g}$ and $\omega_{\alpha g}$ transitions.}
	\label{fig:fig4}
\end{figure*}

Time-resolved studies\cite{Taylor2009} on pentacene thin films have reported weaker featureless ESA contributions on either side of the $S_0 \rightarrow S_1$ GSB signal. The negative ESA features correspond to optically allowed $T_1 \rightarrow T_n$ transitions polarized along the long axis\cite{Tempelaar2017_2} of the molecule, compared to the short axis polarized singlet transitions. The relative oscillator strength and energy shift of the broad and featureless $T_1 \rightarrow T_n$ transitions are not precisely determined, with previous estimates\cite{Tempelaar2017_2,Rao2016} based on fitting simulations to experimental data. For instance, ESA oscillator strengths $|\vec{\mu}_{T_1 \rightarrow T_n}|^2$, relative to $|\vec{\mu}_{S_0 \rightarrow S_1}|^2$, have ranged from $\sim$2.5$^2$ in case of 2DES simulations reported in Rao et al.\cite{Rao2016} to significantly larger value of 15$^2$ in the 2DES simulations of Tempelaar et.al.\cite{Tempelaar2017_2}. The laser bandwidth used for 2D simulations corresponds to a $\sim$ 10 fs Gaussian pulse (field FWHM) with corresponding FWHM spectral range overlaid on the oscillator strengths in \textcolor{black}{Figure~\ref{fig:fig3}a}. Based on this laser bandwidth, our choice of ESA oscillator strengths of $\sim$3.6$^2$ and red-shift of 4803 cm$^{-1}$ is such that the corresponding signal amplitudes in the simulated 2DES spectra are approximately consistent with the experimentally observed\cite{Taylor2009} ratio of GSB signal to the ESA feature at $\sim$1.7 eV. With $n_{v}$ quanta along each vibrational mode on a given diabatic electronic state, the four-wavemixing response function simulations scale expensively with the 4$^{th}$ power of the basis set size as $(5n_v^4)^4$. This necessitates limitations on the number of eigenvectors that can be probed with desired convergence in the resulting non-linear signals. Our choice of the laser bandwidth (corresponding to a 10 fs pulse duration) is partly dictated by this criterion, such that only the lower energy eigenvectors in the $\epsilon- \alpha-\beta$ manifold are probed with a convergence of 2DES CMs to within 2\%. Further details of 2DES simulations and parameters are presented in \textcolor{black}{Section S3}. The convergence checks have been detailed in \textcolor{black}{Section S5}.\\
 %Experimentally, such a map is generated by collecting the third-order signal field at detection frequencies $\omega_t$, as a function of delay $\tau$ between the two pump pulses and the pump-probe waiting time $T$, followed by a Fourier transform along $\tau$.
 %As mentioned in Section \ref{qplusminuscouple}, only anti-correlated inhomogeneity between excited states affects ensemble dephasing\cite{JonasARPC2018} and has been included in the calculations. As before, Homogeneous broadening through optical decoherence is introduced through temperature-dependent Brownian oscillators derived\cite{Zhao2021} from fits of monomer linear spectra (Section \ref{qplusminuscouple}). The simulation details and parameters are described in Section S5.
Figure~\ref{fig:fig4}a shows the simulated absorptive\cite{JonasARPC2003} 2DES spectrum at $T = 100$ fs which reports changes in sample absorption created by the pump as a function of excitation ($\omega_{\tau}$) and detection ($\omega_t$) frequencies. Prominent GSB features with multiple diagonal peaks and associated cross-peaks can be seen. These arise from the two brightest excitons in the mixed $\epsilon-\alpha-\beta$ manifold probed by the laser spectrum (marked as `1' and `2' in \textcolor{black}{Figure~\ref{fig:fig3}}a). The simulated 2D features are more defined (than those observed experimentally) due to intentional absence of inhomogeneity in the 0--1 optical energy gap in our calculations. Calculations of linear and 2DES spectra including this broadening are shown in \si{Section S6} and do not affect the polarization based 2DES signatures of vibronic enhancement proposed here. As also reported\cite{Rao2016} experimentally by Rao et al., no 2D cross-peak between the bright $\alpha-\beta$ manifold and $\epsilon$ is seen because of negligible bright character in $\epsilon_{0000}$. The broad ESA features along the detection axis arise from probe interactions to and from vibrationally excited $(TT)^1$ manifold. Given the impulsive nature of excitation, coherent wavepackets, with purely vibrational or mixed vibronic origins, contribute to the 2DES data as quantum beats. All such ground or excited electronic state wavepackets interfere together and the resultant quantum dynamics can be monitored along the pump-probe waiting time $T$ as shown in Fig.~\ref{fig:fig4}b. The resultant oscillatory signal contributions can be resolved as 2DES CMs along the coherence axis $\omega_T$ by removing the incoherent population background and Fourier transformation along the waiting time $T$. This is schematically described in Figure~\ref{fig:fig4}\textcolor{black}{b,c} with details of the analysis in \textcolor{black}{Section S.5.2}. CM at a given $\omega_T$ reports the 2D amplitude distribution of quantum beats of frequency $\omega_T$. \\

\begin{figure*}[h!]
	\centering
	\includegraphics[width=5 in]{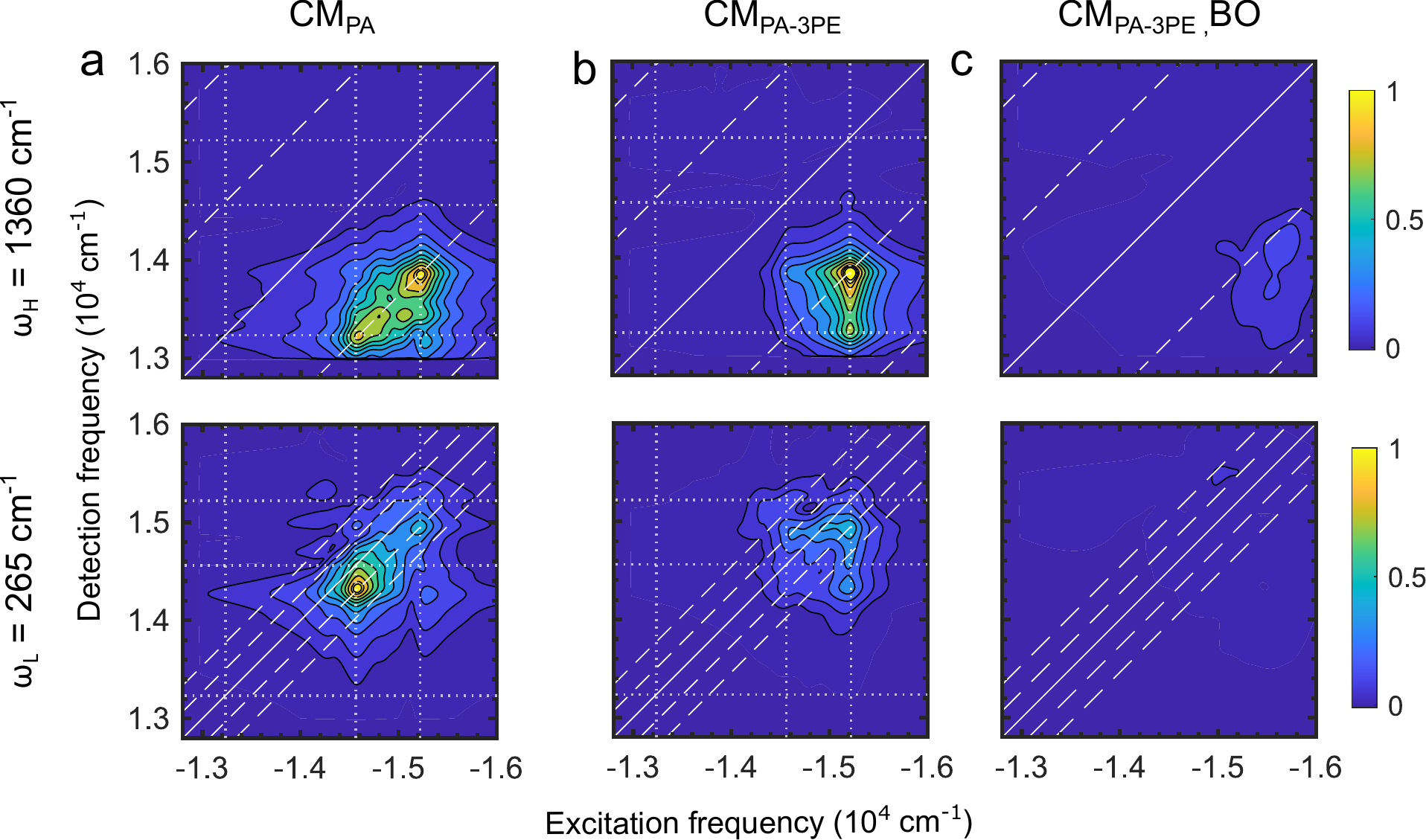}
	\caption{\footnotesize \textbf{(a)} 2DES coherence maps (CMs) for GSB pathways for all-parallel polarization sequence $ZZZZ$ derived from the 2DES maps in Figure~\ref{fig:fig5}. Top and bottom row shows the maps for $\omega_L$ and $\omega_H$, respectively. \textbf{(b)} Corresponding CMs with $PA - 3PE$ polarization sequence. \textbf{(c)} CMs for $PA - 3PE$ sequence but with $\omega_H$ as part of the Brownian oscillator bath such that it does not participate in non-adiabatic mixing through vibronic near-resonance.  Contours are drawn at 5\%,10-90\% at 10\% interval,95\%,100\%. Each map is  normalized with respect to the maxima of the corresponding $CM_{PA}$. Simulations details are described in . Position of the lowest energy dark exciton ($\epsilon_{0000}$) and the brightest two excitons (one exciton for the third panel) under the laser pulse are marked as vertical and horizontal dashed lines. Vertical lines are not shown in panel c because high-frequency mode is not included explicitly in the system Hamiltonian leading to a completely different set of oscillator strengths for this case (see \textcolor{black}{Figure S7}). }	
	\label{fig:fig5}
\end{figure*}

%\vt{HERE}In order to determine whether a given ground state vibrational wavepacket with frequency $\omega_v$, observed in the corresponding CM at $\omega_T = \omega_v$, arises from promoter versus spectator motions, it is instructive to 
%This contribution corresponds to only one of the several terms in the third-order density matrix resulting after three light-matter perturbations,
Following ref.\cite{Jonas2003}, the third order response function expression that leads to a rephasing 2DES GSB signal contribution depends on transition dipole products as 
\footnotesize $\mathcal{R} \sim \langle (\hat{\mu}_{s} \textbf{.} \hat{\mathcal{E}}_s) (\hat{\mu}_{c} \textbf{.} \hat{\mathcal{E}}_c) (\hat{\mu}_{b}\textbf{ .} \hat{\mathcal{E}}_b) (\hat{\mu}_{a} \textbf{.} \hat{\mathcal{E}}_a)\rangle$. 
\normalsize Fields $\hat{\mathcal{E}}_{a-c}$ interact with individual systems to generate the transition dipole $\hat{\mu}_s$. The phase-matched sum of dipole-radiated electric fields (signal field) is detected by interference with a known electric field $\hat{\mathcal{E}}_d$. $\langle\dots\rangle$ denotes isotropic orientational average over the four electric field polarization vectors with respect to the molecular frame transition dipole vectors. The Greens' function time propagator\cite{JonasARPC2003} part has been suppressed for brevity. Using this notation, the 2DES GSB signal contribution that arises from ground state vibrational coherences in a two-electronic level system can be written as \footnotesize $\mathcal{R}_{GSB} \sim \langle (\hat{\mu}_{e_1g_1}\textbf{.}\hat{\mathcal{E}}_s) (\hat{\mu}_{g_0e_1} \textbf{.} \hat{\mathcal{E}}_c) (\hat{\mu}_{g_1e_1} \textbf{.} \hat{\mathcal{E}}_b) (\hat{\mu}_{e_1g_0} \textbf{.} \hat{\mathcal{E}}_a)\rangle$. \normalsize The diagrammatic representation of this contribution as a wavemixing pathway is shown in Figure~\ref{fig:fig5}d. After two interactions, a ground state vibrational coherence between $g_0$ and $g_1$ is created which evolves with $T$. The diagram is explained in detail in \textcolor{black}{Section S4}. Simplifying the transition dipole product for an all-parallel polarization sequence with electric field polarization $\hat{Z}$ in the lab frame, the coherent GSB contribution can be simplified as \footnotesize $\mathcal{R}_{GSB}^{coh} \sim \langle (\hat{\mu}_{ge}\textbf{.}\hat{Z})^4 \rangle |\braket{g_0|e_1}|^2|\braket{g_1|e_1}|^2$\normalsize, where $\hat{\mu}_{ge}$ denotes the transition dipole vector followed by FC overlap factors. The position of this contribution lies at $(\omega_{\tau}, \omega_t) = (\omega_{g_0e_0} + \omega_v, \omega_{g_0e_0} )$, that is, one vibrational quanta above the 2D diagonal along the excitation axis. This can also be deduced from the length of first and last interaction in the wavemixing diagram in Figure \ref{fig:fig5}d. \\
%Even for the simplest case of a two-electronic level system, and ignoring 0--2 FC factors or inter-mode vibrational couplings, all such coherent GSB contributions $\mathcal{R}_{GSB}^{coh}$ interfere together and contribute as a chair pattern\cite{Seibt2013} on the 2D map. The relative magnitude and sign of individual pathways depends on transition dipole strengths and the FC overlap factors.  \\
%In the same fashion, an incoherent rephasing GSB contribution is written as $\mathcal{R}_{GSB}^{incoh} \sim \langle (\hat{\mu}_{ge}\textbf{.}\hat{Z})^4 \rangle |\braket{g_0|e_0}|^4$. 
%This suggests that for a pentacene monomer, when a vibrational wavepacket for the $\omega_H$ is created, a \ab{7.6\%} modulation of the incoherent signal due to vibrational coherences is expected.  This modulation quite substantial compared to the expectation for photosynthetic pigments such as Bacteriochlorophyll a\cite{JonasARPC2018}, and will naturally be readily detectable in a solution or thin film sample. The 2D position of this contribution along excitation $\omega_{\tau}$ and emission $\omega_t$ axes can be inferred by the length of the first and last arrow in the wavemixing diagram in Figure~\ref{fig:fig5}c.

When the $(10)_{\epsilon\alpha}$ near-resonant vibronic manifold is considered (Section \ref{qplusminuscouple}), CMs are further complicated with the wavemixing pathways modified due to vibronic resonance as shown in Figure~\ref{fig:fig5}e. Considering vibronic mixing to dominantly arise from only within the degenerate manifold, the mixed eigenvectors can be approximated\cite{Sahu2020} as $\ket{\psi_{\pm}} \sim (\ket{\epsilon_{10}} \pm \ket{\alpha_{00}})/\sqrt{2}$. Wavemixing pathways arising from these eigenvectors lead to dipole products such as \footnotesize $\langle (\hat{\mu}_{\pm g_1}\textbf{.}\hat{\mathcal{E}}_d) (\hat{\mu}_{g_0\pm} \textbf{.} \hat{\mathcal{E}}_c) (\hat{\mu}_{g_1\pm} \textbf{.} \hat{\mathcal{E}}_b) (\hat{\mu}_{\pm g_0} \textbf{.} \hat{\mathcal{E}}_a)\rangle$ \normalsize in the response function. All the pathways resulting from this dipole product contribute at the 2D cross-peak location between the negligible lower diagonal 2D peak (DP$_L$, arising from direct $g \rightarrow \epsilon$ excitation), and the bright upper diagonal 2D peak (DP$_U$, arising from direct $g \rightarrow \alpha$ excitation). Expansion of the dipole product shows that even without any vibronic mixing from the bright exciton $\alpha$, a wavemixing pathway equivalent to Figure~\ref{fig:fig5}d will arise, however with a negligible contribution due to its $|\mu_{g \epsilon}|^4$ dependence ($\epsilon$ contains only ~7\% bright character, \textcolor{black}{see Table S5}). Similarly, dominant terms dependent on $|\mu_{g \alpha}|^4$ will also contribute at the same 2D cross peak location. However, it is important to recognize that such terms are not specific to vibronic mixing because they can arise even without any $\epsilon-\alpha$ vibronic mixing.  Jonas and co-workers have shown\cite{Tiwari2013} that non-adiabatic mixing due to vibronic resonance on the excited state leads to enhancement ground state vibrational quantum beats at the 2DES cross peak in case of photosynthetic excitons. Such beats are otherwise weak because small HR factors in photosynthetic pigments render wavemixing pathways such as those in Figure \ref{fig:fig4}d  very weak on account of poor $|\braket{g_0|e_1}|^2$ FC overlaps. The same is \textit{not} true in the SEF context however. Large HR factors in acenes imply that dominant beating amplitude at the cross peak can readily arise from wavemixing pathways with non-specific dipole products such as $|\mu_{g\alpha}|^4$ in Figure \ref{fig:fig4}e. Wavemixing pathways starting from vibrationally hot ground state, allowed in the simulations, further complicate the interpretation of quantum beating signals. Thus, \textit{any inferences regarding vibronic enhancement of SEF cannot be made on the basis of readily detectable quantum beat amplitude at the 2DES cross-peak location.} \\
%Figure S9 shows the wavemixing diagrams corresponding to DP$_{L,U}$. 

%\subsection{Unique 2DES Spectroscopic Signatures of Vibronic Mixing in SEF}\label{deltacm}
\FloatBarrier

A distinction between vibronic mixing dependent pathways can be motivated by considering the dipole product cross-terms in the wavemixing pathway in Figure \ref{fig:fig5}e. Cross-terms such as $|\mu_{g\alpha}|^2|\mu_{g\epsilon}|^2$ only arise in the presence of vibronic mixing. To motivate a polarization-based signature of quantum beats arising due to vibronic mixing akin to vibrational polarization anisotropy\cite{Jonas2008,Jonas2008b}, we will first denote dipole products such as \footnotesize$\langle (\hat{\mu}_{\epsilon_1 g_1}\textbf{.}\hat{\mathcal{E}}_d) (\hat{\mu}_{g_0\epsilon_1} \textbf{.} \hat{\mathcal{E}}_c) (\hat{\mu}_{g_1\epsilon_1} \textbf{.} \hat{\mathcal{E}}_b) (\hat{\mu}_{\epsilon_1 g_0} \textbf{.} \hat{\mathcal{E}}_a)\rangle$ \normalsize in a compact notation as, $\langle \epsilon_Z \epsilon_Z \epsilon_Z \epsilon_Z\rangle$ for an all-parallel electric field polarization $\hat{Z}$ in the lab frame. Alternatively, considering a perpendicular pump-probe polarization sequence, the dipole product will be denoted as  $\langle \epsilon_Z \epsilon_Z \epsilon_Y \epsilon_Y\rangle$. The other non-specific dipole product is denoted as $\langle \alpha_Z \alpha_Z \alpha_Y \alpha_Y\rangle$. The isotropic orientational averages for the parallel ($PA$, $ZZZZ$) and perpendicular ($PE$, $ZZYY$) sequences are related\cite{Jonas2003} by a factor of 1/3, implying that a 2DES CM evaluated at a mixed polarization sequence $PA - 3PE$ will cancel out the non-specific ground state quantum beat pathways because they are isotropic. In contrast, the vibronic mixing specific cross-terms such as $\langle \alpha_Z \epsilon_Z \alpha_Y \epsilon_Y\rangle$ in Figure \ref{fig:fig5}e are anisotropic and survive the $PA - 3PE$ sequence. These will only arise in the presence of  vibronic mixing without any masking from non-specific vibronic mixing contributions. The wavemixing diagram for the most dominant surviving pathway is shown in Figure~\ref{fig:fig5}e where interactions utilizing $\hat{\mu}_{g\epsilon}$ and $\hat{\mu}_{g\alpha}$ transition dipole vectors are shown in different color to denote their different directions. All other surviving pathways are analyzed in \textcolor{black}{Section S4}. 

The above polarization sequence to isolate vibronic mixing specific terms assumes a large laser focal spot, typical of conventional 2DES approaches\cite{Fuller2015}, to sample an orientationally averaged response of the thin film. Below we apply this idea to the model SEF Hamiltonian to spectroscopically confirm the synergistic role of low- and high-frequency vibrations in vibronic enhancement of SEF. \vt{In order to isolate the GSB wavemixing pathways in Fig.~\ref{fig:fig4}d which may arise due to vibronic enhancement, the laser spectrum should cover the lowest energy triplet exciton as well as the low-energy bright excitons ($g \rightarrow \epsilon$ and $g \rightarrow \alpha$ in the pentacene dimer). This can be understood by considering the vertical lengths of interactions in Fig.~\ref{fig:fig4}d. The laser spectrum in Fig.~\ref{fig:fig3}a is chosen accordingly.} Such an experiment can be readily implemented to test these predictions and identify which, if any, vibrational modes promote vibronically enhanced SEF with associated enhancement of vibrational quantum beats monitored through spectroscopy.
%The $|\mu_{g \epsilon}|^4$ and $|\mu_{g \alpha}|^4$ dependence in GSB quantum beat contributions arises from dipole products such as --
%\begin{eqnarray}
%	\langle (\hat{\mu}_{\epsilon_1 g_1}\textbf{.}\hat{\mathcal{E}}_d) (\hat{\mu}_{g_0\epsilon_1} \textbf{.} \hat{\mathcal{E}}_c) (\hat{\mu}_{g_1\epsilon_1} \textbf{.} \hat{\mathcal{E}}_b) (\hat{\mu}_{\epsilon_1 g_0} \textbf{.} \hat{\mathcal{E}}_a)\rangle &\sim& |\mu_{g \epsilon}|^4 .|\braket{1^\epsilon|0^g}|^2|\braket{1^\epsilon|1^g}|^2 \nonumber \\
%	\langle (\hat{\mu}_{\alpha_0 g_1}\textbf{.}\hat{\mathcal{E}}_d) (\hat{\mu}_{g_0\alpha_0} \textbf{.} \hat{\mathcal{E}}_c) (\hat{\mu}_{g_1\alpha_0} \textbf{.} \hat{\mathcal{E}}_b) (\hat{\mu}_{\alpha_0 g_0} \textbf{.} \hat{\mathcal{E}}_a)\rangle &\sim& |\mu_{g \alpha}|^4.|\braket{0^\alpha|1^g}|^2|\braket{0^\alpha|0^g}|^2\nonumber
%\end{eqnarray}
%Such dipole products contribute at the 2DES cross-peak and arise with or without $\epsilon-\alpha$ vibronic mixing, and therefore not specific to vibronic mixing. The above dipole products will be compactly denoted as $\langle \epsilon_Z \epsilon_Z \epsilon_Z \epsilon_Z\rangle$ and $\langle \alpha_Z \alpha_Z \alpha_Z \alpha_Z\rangle$, respectively, for an all-parallel electric field polarization sequence. 

%Further, the role of laser bandwidth position is crucial to eliminate \vt{HERE - mention laser bandwidth crucial and asssumed orientational average and other more complicated schemes} 

Figure~\ref{fig:fig5} shows 2DES CMs at $\omega_T = \omega_L$ and $\omega_T = \omega_H$. Figure~\ref{fig:fig5}a shows 2DES CMs for the Hamiltonian in Equation~\ref{eq1} where both $\omega_H$ and $\omega_L$ intramolecular vibrations are included explicitly in the system Hamiltonian and therefore allowed to participate in vibronic mixing. The map is generated by an all-parallel polarization sequence $ZZZZ$ and therefore includes wavemixing pathways which are both specific as well as non-specific to vibronic mixing as explained in Figure~\ref{fig:fig5}d and  Figure~\ref{fig:fig5}e. The beating amplitude appears at the 2D cross-peak location between $\omega_{g \epsilon}$ and $\omega_{g \alpha/\beta}$ diagonal peaks, even though the former is not seen\cite{Rao2016} in the 2DES map in Figure~\ref{fig:fig5}a. This is so because a $|\mu_{\epsilon g}|^4$ dipole product dependence renders it negligibly weak. The CMs do however show the oscillatory cross-peak contributions as being quite prominent. The beating amplitude appears at expected location of approximately one $\omega_H$ vibrational quanta below the diagonal along the detection axis. Similar features have been experimentally reported\cite{Rao2016} and interpreted as signatures of vibronic mixing. However, as explained earlier in this section, it would be misleading to interpret such quantum beats as vibronically enhanced solely based on their 2D location and prominent amplitude because of dominant non-specific vibronic mixing contributions, such as $\langle \alpha_Z \alpha_Z \alpha_Z \alpha_Z\rangle$, at the same 2D cross peak location. For the same reason, the CMs for $\omega_H$ and $\omega_L$ cannot confirm the prediction in Section \ref{peierls} that Peierls coupling by itself, without vibronic near-resonance through the high-frequency vibration, is ineffective in promoting vibronic mixing (see Fig.\ref{fig:fig2}). 2DES signals arising from vibrationally hot ground states further complicate the CMs. \\

Next we apply the $PA-3PE$ polarization sequence to eliminate non-specific isotropic quantum beat contributions. Figure~\ref{fig:fig5}b plots the resulting CMs for both $\omega_H$ and $\omega_L$. As expected for $\omega_H$, cross dipole product contributions such as $\langle \alpha_Z \epsilon_Z \alpha_Z \epsilon_Z\rangle$ and $\langle \alpha_Z \epsilon_Z \alpha_Y \epsilon_Y\rangle$ survive the $PA-3PE$ sequence to result in the CMs in panel b. Interestingly, as also expected, $\omega_L$ CM contributions also survive because their complementary role in near-resonant vibronic mixing leads to mixed polarization signatures for $\omega_L$ quantum beats as well. To further confirm the synergistic effect of low-frequency vibrations on vibronic near-resonance, we remove the $\omega_H$ mode from the system Hamiltonian and include it in the Brownian oscillator bath while keeping its total stabilization energy same as before. The low-frequency FC active vibration as well as its Peierls coupling is still treated explicitly in the system Hamiltonian as before. The resulting CMs are shown in panel c. Since $\omega_H$ is not treated explicitly in the system, it can no longer participate in non-adiabatic vibronic mixing. Consequently, as may be expected, its CM does not survive the $PA-3PE$ sequence. Additionally, the $\omega_L$ CM also does not survive the $PA-3PE$ sequence confirming the \textit{complementary enhancement of vibronic mixing through low-frequency modulations of intermolecular orbital overlaps when a vibronic near-resonance due to high-frequency vibrations is expected}. 

It can be easily shown that experimentally one does not need to perform two different experiments, one each with $PA$ and $PE$ sequence. Rather a single experiment with polarization sequence $(0^o0^o+60^o-60^o)$ leads to a scaled $PA-3PE$ signal. Such pulse polarization sequences were originally devised\cite{Zanni2001} by Zanni and Hochstrasser in the context of eliminating 2D diagonal peaks in 2DIR spectra. More complex polarization schemes specific to vibronic mixing have also been implemented\cite{Thyrhaug2018}. However the $(0^o0^o+60^o-60^o)$ sequence is fairly easy to implement compared to other schemes. Zanni and co-workers have recently\cite{Zanni2022} extended this sequence to incorporate rotational diffusion effects and isolate 2D cross-peak in a \textit{pump-probe} experiment by eliminating diagonal peak contributions. Here we have extended this concept in the context of polarization dependence of vibronic quantum beats to isolate anisotropic quantum beat wavemixing pathways that are \textit{specific} to vibronic mixing in SEF. 

\section{Conclusions}
Resonances between electronic energy gaps in photosynthetic proteins and intramolecular vibrational frequencies of the pigments leads to a non-adiabatic energy funnel\cite{JonasARPC2018} driven by anti-phase vibrational motions between the two pigments. Such resonances are narrow and therefore not robust to energetic mismatches. In-phase motions between the two molecules are mere spectators\cite{Moffit1960, Gouterman1961, Tiwari2013}  in this process. Similar effects have been reported\cite{Rao2016,Tempelaar2018_3,Tempelaar2022} in case of SEF. We have introduced fresh, and previously overlooked, mechanistic insights into this emerging understanding of the rapid formation of  $(TT)^1$ state. We show that, quite counterintuitively, in-phase vibrational motions between the two pentacene molecules drive non-adiabatic vibronic mixing between the bright exciton and the optically dark $(TT)^1$ manifold, while anti-phase vibrational motions mix the optically bright excitons together, as in an excitonic dimer. Together these vibrational motions lead to a coupled vibronic resonances between different pairs of electronic states, driven by a single high-frequency vibration on each molecule. Significantly, this occurs over a broad non-selective range of vibrational frequencies and is in striking contrast to that in photosynthetic excitons. 

Modulations of intermolecular orbital overlaps in pentacene thin films are expected from known\cite{Venuti2002,Venuti2004} low-frequency vibrations with mixed inter/intramolecular character. Significantly, introducing such a vibration in the above picture leads to an overall synergistic effect of low- and high-frequency vibrations in SEF. The existing vibronic near-resonances become more robust to energetic mismatches with increased vibronic exciton delocalization, and consequently lead to increased formation of $(TT)^1$ upon impulsive photoexcitation of bright electronic states. Treatment of low-frequency vibrations beyond a `frozen' mode\cite{Tempelaar2018_3} leads to these interesting effects and resolves conflicting\cite{Tempelaar2018_3,Huo2017} and often exclusive\cite{Duan2020} roles of vibronic resonance and Peierls' coupling in enhancing SEF.  

Quantum beats in 2DES often carry vital mechanistic insights\cite{JonasARPC2018} into energy and charge transfer. In SEF, prominent vibrational quantum beats are already expected in pentacene monomers because of large HR factors. We argue that readily observed\cite{Rao2016, Duan2020} quantum beat amplitudes at 2DES at cross-peak locations therefore cannot unambiguously identify vibronic mixing effects in SEF. Instead, by analyzing the polarization anisotropy of vibrational quantum beats in the clean ground state signals available in SEF, we propose and demonstrate polarization-based 2DES signatures that survive only in the presence of excited state vibronic mixing. Such a polarization scheme is readily implementable, can distinguish vibrational quantum beats that arise from vibronic mixing and promote SEF against those which merely accompany the process as spectator modes, and makes the predictions of our model amenable to experimental scrutiny.

Multiple electronic states, explicit quantum treatment of multiple vibrational modes, and no approximation\cite{Rao2016} of ground state vibrations in vibronic resonance are together vital in order to capture the  previously unrecognized complementary roles of vibronic coupling from low- and high-frequency vibrations, as well as of in-phase and anti-phase vibrational motions. Even though our calculations, with experimentally relevant parameters, are based on a pentacene dimer and do not capture the thermodynamic enhancement\cite{Tempelaar2018_3} of SEF rates possible in a crystal, the physical insights about robust, non-selective and coupled vibronic resonances may have significant implications even for a pentacene thin film. For example, coupled resonances, but along a high- and low-frequency mode, seem likely to occur within a Davydov component given its $\sim$ 1200 cm$^{-1}$ width\cite{Hestand2015}. Similarly, experimentally known energetic separation of $\sim 1000$ cm$^{-1}$ between upper and lower Davydov components in pentacene thin films\cite{Hestand2015}  is of the order of a high-frequency vibrational excitation. Ultrafast decay of electronic polarization anisotropy\cite{Pensack2015} in pentacene nanoparticles on the timescale of $(TT)^1$ formation is also suggestive of highly mixed vibronic states at play. Physical insights gained from our model are more general in nature and may be applicable to intramolecular SEF systems as well. Our analysis motivates synthetic design principles that go beyond tuning electronic couplings\cite{Patil2019} to incorporate high-frequency intramolecular vibrations with large Huang-Rhys factors, exciton energy gaps of the order of high frequency vibrations, and low-frequency vibrations capable of modulating intermolecular orbital overlaps.

\bibliographystyle{unsrt}
\bibliography{SEFrefs}

\providecommand{\latin}[1]{#1}
\makeatletter
\providecommand{\doi}
  {\begingroup\let\do\@makeother\dospecials
  \catcode`\{=1 \catcode`\}=2 \doi@aux}
\providecommand{\doi@aux}[1]{\endgroup\texttt{#1}}
\makeatother
\providecommand*\mcitethebibliography{\thebibliography}
\csname @ifundefined\endcsname{endmcitethebibliography}
  {\let\endmcitethebibliography\endthebibliography}{}
\begin{mcitethebibliography}{80}
\providecommand*\natexlab[1]{#1}
\providecommand*\mciteSetBstSublistMode[1]{}
\providecommand*\mciteSetBstMaxWidthForm[2]{}
\providecommand*\mciteBstWouldAddEndPuncttrue
  {\def\EndOfBibitem{\unskip.}}
\providecommand*\mciteBstWouldAddEndPunctfalse
  {\let\EndOfBibitem\relax}
\providecommand*\mciteSetBstMidEndSepPunct[3]{}
\providecommand*\mciteSetBstSublistLabelBeginEnd[3]{}
\providecommand*\EndOfBibitem{}
\mciteSetBstSublistMode{f}
\mciteSetBstMaxWidthForm{subitem}{(\alph{mcitesubitemcount})}
\mciteSetBstSublistLabelBeginEnd
  {\mcitemaxwidthsubitemform\space}
  {\relax}
  {\relax}

\bibitem[Smith and Michl(2013)Smith, and Michl]{Michl2013}
Smith,~M.~B.; Michl,~J. {Recent Advances in Singlet Fission}. \emph{Annual
  Review of Physical Chemistry} \textbf{2013}, \emph{64}, 361--386\relax
\mciteBstWouldAddEndPuncttrue
\mciteSetBstMidEndSepPunct{\mcitedefaultmidpunct}
{\mcitedefaultendpunct}{\mcitedefaultseppunct}\relax
\EndOfBibitem
\bibitem[Zimmerman \latin{et~al.}(2011)Zimmerman, Bell, Casanova, and
  Head-Gordon]{Zimmerman2011}
Zimmerman,~P.~M.; Bell,~F.; Casanova,~D.; Head-Gordon,~M. {Mechanism for
  Singlet Fission in Pentacene and Tetracene: From Single Exciton to Two
  Triplets}. \emph{Journal of the American Chemical Society} \textbf{2011},
  \emph{133}, 19944--19952\relax
\mciteBstWouldAddEndPuncttrue
\mciteSetBstMidEndSepPunct{\mcitedefaultmidpunct}
{\mcitedefaultendpunct}{\mcitedefaultseppunct}\relax
\EndOfBibitem
\bibitem[Zeng \latin{et~al.}(2014)Zeng, Hoffmann, and Ananth]{Ananth2014}
Zeng,~T.; Hoffmann,~R.; Ananth,~N. {The Low-Lying Electronic States of
  Pentacene and Their Roles in Singlet Fission}. \emph{Journal of the American
  Chemical Society} \textbf{2014}, \emph{136}, 5755--5764\relax
\mciteBstWouldAddEndPuncttrue
\mciteSetBstMidEndSepPunct{\mcitedefaultmidpunct}
{\mcitedefaultendpunct}{\mcitedefaultseppunct}\relax
\EndOfBibitem
\bibitem[Feng \latin{et~al.}(2013)Feng, Luzanov, and Krylov]{Krylov2013}
Feng,~X.; Luzanov,~A.~V.; Krylov,~A.~I. {Fission of Entangled Spins: An
  Electronic Structure Perspective}. \emph{The Journal of Physical Chemistry
  Letters} \textbf{2013}, \emph{4}, 3845--3852\relax
\mciteBstWouldAddEndPuncttrue
\mciteSetBstMidEndSepPunct{\mcitedefaultmidpunct}
{\mcitedefaultendpunct}{\mcitedefaultseppunct}\relax
\EndOfBibitem
\bibitem[Phys \latin{et~al.}(2014)Phys, Berkelbach, Hybertsen, and
  Reichman]{Berkelbach2}
Phys,~J.~C.; Berkelbach,~T.~C.; Hybertsen,~M.~S.; Reichman,~D.~R. {Microscopic
  theory of singlet exciton fission . II . Application to pentacene dimers and
  the role of superexchange}. \textbf{2014}, \emph{114103}\relax
\mciteBstWouldAddEndPuncttrue
\mciteSetBstMidEndSepPunct{\mcitedefaultmidpunct}
{\mcitedefaultendpunct}{\mcitedefaultseppunct}\relax
\EndOfBibitem
\bibitem[Jones \latin{et~al.}(2020)Jones, Kearns, Ho, Flach, and
  Zanni]{Jones2020}
Jones,~A.~C.; Kearns,~N.~M.; Ho,~J.-J.; Flach,~J.~T.; Zanni,~M.~T. Impact of
  non-equilibrium molecular packings on singlet fission in microcrystals
  observed using 2D white-light microscopy. \emph{Nature Chemistry}
  \textbf{2020}, \emph{12}, 40--47\relax
\mciteBstWouldAddEndPuncttrue
\mciteSetBstMidEndSepPunct{\mcitedefaultmidpunct}
{\mcitedefaultendpunct}{\mcitedefaultseppunct}\relax
\EndOfBibitem
\bibitem[Tempelaar and Reichman(2017)Tempelaar, and Reichman]{Tempelaar2017_1}
Tempelaar,~R.; Reichman,~D.~R. {Vibronic exciton theory of singlet fission. I.
  Linear absorption and the anatomy of the correlated triplet pair state}.
  \emph{The Journal of Chemical Physics} \textbf{2017}, \emph{146},
  174703\relax
\mciteBstWouldAddEndPuncttrue
\mciteSetBstMidEndSepPunct{\mcitedefaultmidpunct}
{\mcitedefaultendpunct}{\mcitedefaultseppunct}\relax
\EndOfBibitem
\bibitem[Tempelaar and Reichman(2017)Tempelaar, and Reichman]{Tempelaar2017_2}
Tempelaar,~R.; Reichman,~D.~R. {Vibronic exciton theory of singlet fission. II.
  Two-dimensional spectroscopic detection of the correlated triplet pair
  state}. \emph{The Journal of Chemical Physics} \textbf{2017}, \emph{146},
  174704\relax
\mciteBstWouldAddEndPuncttrue
\mciteSetBstMidEndSepPunct{\mcitedefaultmidpunct}
{\mcitedefaultendpunct}{\mcitedefaultseppunct}\relax
\EndOfBibitem
\bibitem[Tempelaar and Reichman(2018)Tempelaar, and Reichman]{Tempelaar2018_3}
Tempelaar,~R.; Reichman,~D.~R. {Vibronic exciton theory of singlet fission.
  III. How vibronic coupling and thermodynamics promote rapid triplet
  generation in pentacene crystals}. \emph{The Journal of Chemical Physics}
  \textbf{2018}, \emph{148}, 244701\relax
\mciteBstWouldAddEndPuncttrue
\mciteSetBstMidEndSepPunct{\mcitedefaultmidpunct}
{\mcitedefaultendpunct}{\mcitedefaultseppunct}\relax
\EndOfBibitem
\bibitem[Unger \latin{et~al.}(2022)Unger, Moretti, Hausch, Bredehoeft, Zeiser,
  Haug, Tempelaar, Hestand, Cerullo, and Broch]{Tempelaar2022}
Unger,~F.; Moretti,~L.; Hausch,~J.; Bredehoeft,~J.; Zeiser,~C.; Haug,~S.;
  Tempelaar,~R.; Hestand,~N.~J.; Cerullo,~G.; Broch,~K. {Modulating Singlet
  Fission by Scanning through Vibronic Resonances in Pentacene-Based Blends}.
  \emph{Journal of the American Chemical Society} \textbf{2022}, \emph{144},
  20610--20619\relax
\mciteBstWouldAddEndPuncttrue
\mciteSetBstMidEndSepPunct{\mcitedefaultmidpunct}
{\mcitedefaultendpunct}{\mcitedefaultseppunct}\relax
\EndOfBibitem
\bibitem[Morrison and Herbert(2017)Morrison, and Herbert]{Herbert2017}
Morrison,~A.~F.; Herbert,~J.~M. Evidence for Singlet Fission Driven by Vibronic
  Coherence in Crystalline Tetracene. \emph{The Journal of Physical Chemistry
  Letters} \textbf{2017}, \emph{8}, 1442--1448, PMID: 28277682\relax
\mciteBstWouldAddEndPuncttrue
\mciteSetBstMidEndSepPunct{\mcitedefaultmidpunct}
{\mcitedefaultendpunct}{\mcitedefaultseppunct}\relax
\EndOfBibitem
\bibitem[Duan \latin{et~al.}(2020)Duan, Jha, Li, Tiwari, Ye, Nayak, Zhu, Li,
  Martinez, Thorwart, and Miller]{Duan2020}
Duan,~H.-G.; Jha,~A.; Li,~X.; Tiwari,~V.; Ye,~H.; Nayak,~P.~K.; Zhu,~X.-L.;
  Li,~Z.; Martinez,~T.~J.; Thorwart,~M.; Miller,~R. J.~D. {Intermolecular
  vibrations mediate ultrafast singlet fission}. \emph{Science Advances}
  \textbf{2020}, \emph{6}\relax
\mciteBstWouldAddEndPuncttrue
\mciteSetBstMidEndSepPunct{\mcitedefaultmidpunct}
{\mcitedefaultendpunct}{\mcitedefaultseppunct}\relax
\EndOfBibitem
\bibitem[Bakulin \latin{et~al.}(2016)Bakulin, Morgan, Kehoe, Wilson, Chin,
  Zigmantas, Egorova, and Rao]{Rao2016}
Bakulin,~A.~A.; Morgan,~S.~E.; Kehoe,~T.~B.; Wilson,~M. W.~B.; Chin,~A.~W.;
  Zigmantas,~D.; Egorova,~D.; Rao,~A. {Real-time observation of multiexcitonic
  states in ultrafast singlet fission using coherent 2D electronic
  spectroscopy}. \emph{Nature Chemistry} \textbf{2016}, \emph{8}, 16--23\relax
\mciteBstWouldAddEndPuncttrue
\mciteSetBstMidEndSepPunct{\mcitedefaultmidpunct}
{\mcitedefaultendpunct}{\mcitedefaultseppunct}\relax
\EndOfBibitem
\bibitem[Womick and Moran(2011)Womick, and Moran]{Womick2011}
Womick,~J.~M.; Moran,~A.~M. {Vibronic Enhancement of Exciton Sizes and Energy
  Transport in Photosynthetic Complexes}. \emph{J. Phys. Chem. B}
  \textbf{2011}, \emph{115}, 1347--1356\relax
\mciteBstWouldAddEndPuncttrue
\mciteSetBstMidEndSepPunct{\mcitedefaultmidpunct}
{\mcitedefaultendpunct}{\mcitedefaultseppunct}\relax
\EndOfBibitem
\bibitem[Tiwari \latin{et~al.}(2013)Tiwari, Peters, and Jonas]{Tiwari2013}
Tiwari,~V.; Peters,~W.~K.; Jonas,~D.~M. {Electronic resonance with
  anticorrelated pigment vibrations drives photosynthetic energy transfer
  outside the adiabatic framework}. \emph{Proceedings of the National Academy
  of Sciences} \textbf{2013}, \emph{110}, 1203--1208\relax
\mciteBstWouldAddEndPuncttrue
\mciteSetBstMidEndSepPunct{\mcitedefaultmidpunct}
{\mcitedefaultendpunct}{\mcitedefaultseppunct}\relax
\EndOfBibitem
\bibitem[Peters \latin{et~al.}(2017)Peters, Tiwari, and Jonas]{Peters2017}
Peters,~W.~K.; Tiwari,~V.; Jonas,~D.~M. {Nodeless vibrational amplitudes and
  quantum nonadiabatic dynamics in the nested funnel for a pseudo Jahn-Teller
  molecule or homodimer}. \emph{The Journal of Chemical Physics} \textbf{2017},
  \emph{147}, 194306\relax
\mciteBstWouldAddEndPuncttrue
\mciteSetBstMidEndSepPunct{\mcitedefaultmidpunct}
{\mcitedefaultendpunct}{\mcitedefaultseppunct}\relax
\EndOfBibitem
\bibitem[Jonas(2018)]{JonasARPC2018}
Jonas,~D.~M. {Vibrational and Nonadiabatic Coherence in 2D Electronic
  Spectroscopy, the Jahn–Teller Effect, and Energy Transfer}. \emph{Annual
  Review of Physical Chemistry} \textbf{2018}, \emph{69}, 327--352\relax
\mciteBstWouldAddEndPuncttrue
\mciteSetBstMidEndSepPunct{\mcitedefaultmidpunct}
{\mcitedefaultendpunct}{\mcitedefaultseppunct}\relax
\EndOfBibitem
\bibitem[Tiwari \latin{et~al.}(2017)Tiwari, Peters, and Jonas]{Tiwari2017}
Tiwari,~V.; Peters,~W.~K.; Jonas,~D.~M. {Electronic energy transfer through
  non-adiabatic vibrational-electronic resonance. I. Theory for a dimer}.
  \emph{The Journal of Chemical Physics} \textbf{2017}, \emph{147},
  154308\relax
\mciteBstWouldAddEndPuncttrue
\mciteSetBstMidEndSepPunct{\mcitedefaultmidpunct}
{\mcitedefaultendpunct}{\mcitedefaultseppunct}\relax
\EndOfBibitem
\bibitem[Tiwari and Jonas(2018)Tiwari, and Jonas]{Tiwari2018}
Tiwari,~V.; Jonas,~D.~M. {Electronic energy transfer through non-adiabatic
  vibrational-electronic resonance. II. 1D spectra for a dimer}. \emph{The
  Journal of Chemical Physics} \textbf{2018}, \emph{148}, 84308\relax
\mciteBstWouldAddEndPuncttrue
\mciteSetBstMidEndSepPunct{\mcitedefaultmidpunct}
{\mcitedefaultendpunct}{\mcitedefaultseppunct}\relax
\EndOfBibitem
\bibitem[Brillante \latin{et~al.}(2002)Brillante, {Della Valle}, Farina,
  Girlando, Masino, and Venuti]{Venuti2002}
Brillante,~A.; {Della Valle},~R.~G.; Farina,~L.; Girlando,~A.; Masino,~M.;
  Venuti,~E. {Raman phonon spectra of pentacene polymorphs}. \emph{Chemical
  Physics Letters} \textbf{2002}, \emph{357}, 32--36\relax
\mciteBstWouldAddEndPuncttrue
\mciteSetBstMidEndSepPunct{\mcitedefaultmidpunct}
{\mcitedefaultendpunct}{\mcitedefaultseppunct}\relax
\EndOfBibitem
\bibitem[{Della Valle} \latin{et~al.}(2004){Della Valle}, Venuti, Farina,
  Brillante, Masino, and Girlando]{Venuti2004}
{Della Valle},~R.~G.; Venuti,~E.; Farina,~L.; Brillante,~A.; Masino,~M.;
  Girlando,~A. {Intramolecular and Low-Frequency Intermolecular Vibrations of
  Pentacene Polymorphs as a Function of Temperature}. \emph{The Journal of
  Physical Chemistry B} \textbf{2004}, \emph{108}, 1822--1826\relax
\mciteBstWouldAddEndPuncttrue
\mciteSetBstMidEndSepPunct{\mcitedefaultmidpunct}
{\mcitedefaultendpunct}{\mcitedefaultseppunct}\relax
\EndOfBibitem
\bibitem[Le \latin{et~al.}(2021)Le, de~la Perrelle, Do, Leng, Tapping, Scholes,
  Kee, and Tan]{Tan2021}
Le,~D.~V.; de~la Perrelle,~J.~M.; Do,~T.~N.; Leng,~X.; Tapping,~P.~C.;
  Scholes,~G.~D.; Kee,~T.~W.; Tan,~H.-S. {Characterization of the ultrafast
  spectral diffusion and vibronic coherence of TIPS-pentacene using 2D
  electronic spectroscopy}. \emph{The Journal of Chemical Physics}
  \textbf{2021}, \emph{155}, 14302\relax
\mciteBstWouldAddEndPuncttrue
\mciteSetBstMidEndSepPunct{\mcitedefaultmidpunct}
{\mcitedefaultendpunct}{\mcitedefaultseppunct}\relax
\EndOfBibitem
\bibitem[Zhu \latin{et~al.}(1994)Zhu, Sage, and Champion]{Champion1994}
Zhu,~L.; Sage,~J.~T.; Champion,~P.~M. {Observation of Coherent Reaction
  Dynamics in Heme Proteins}. \emph{Science} \textbf{1994}, \emph{266},
  629--632\relax
\mciteBstWouldAddEndPuncttrue
\mciteSetBstMidEndSepPunct{\mcitedefaultmidpunct}
{\mcitedefaultendpunct}{\mcitedefaultseppunct}\relax
\EndOfBibitem
\bibitem[Wynne \latin{et~al.}(1996)Wynne, Reid, and Hochstrasser]{Wynne1996}
Wynne,~K.; Reid,~G.~D.; Hochstrasser,~R.~M. {Vibrational coherence in electron
  transfer: The tetracyanoethylene–pyrene complex}. \emph{The Journal of
  Chemical Physics} \textbf{1996}, \emph{105}, 2287--2297\relax
\mciteBstWouldAddEndPuncttrue
\mciteSetBstMidEndSepPunct{\mcitedefaultmidpunct}
{\mcitedefaultendpunct}{\mcitedefaultseppunct}\relax
\EndOfBibitem
\bibitem[Wolfseder \latin{et~al.}(1998)Wolfseder, Seidner, Domcke, Stock, Seel,
  Engleitner, and Zinth]{Zinth1998}
Wolfseder,~B.; Seidner,~L.; Domcke,~W.; Stock,~G.; Seel,~M.; Engleitner,~S.;
  Zinth,~W. {Vibrational coherence in ultrafast electron-transfer dynamics of
  oxazine 1 in N,N-dimethylaniline: simulation of a femtosecond pump-probe
  experiment}. \emph{Chemical Physics} \textbf{1998}, \emph{233},
  323--334\relax
\mciteBstWouldAddEndPuncttrue
\mciteSetBstMidEndSepPunct{\mcitedefaultmidpunct}
{\mcitedefaultendpunct}{\mcitedefaultseppunct}\relax
\EndOfBibitem
\bibitem[Kim \latin{et~al.}(2012)Kim, Kim, Park, Ko, Lee, and Joo]{Joo2012}
Kim,~S.~Y.; Kim,~C.~H.; Park,~M.; Ko,~K.~C.; Lee,~J.~Y.; Joo,~T. {Coherent
  Nuclear Wave Packets Generated by Ultrafast Intramolecular Charge-Transfer
  Reaction}. \emph{The Journal of Physical Chemistry Letters} \textbf{2012},
  \emph{3}, 2761--2766\relax
\mciteBstWouldAddEndPuncttrue
\mciteSetBstMidEndSepPunct{\mcitedefaultmidpunct}
{\mcitedefaultendpunct}{\mcitedefaultseppunct}\relax
\EndOfBibitem
\bibitem[Rafiq \latin{et~al.}(2021)Rafiq, Fu, Kudisch, and
  Scholes]{Scholes2021}
Rafiq,~S.; Fu,~B.; Kudisch,~B.; Scholes,~G.~D. {Interplay of vibrational
  wavepackets during an ultrafast electron transfer reaction}. \emph{Nature
  Chemistry} \textbf{2021}, \emph{13}, 70--76\relax
\mciteBstWouldAddEndPuncttrue
\mciteSetBstMidEndSepPunct{\mcitedefaultmidpunct}
{\mcitedefaultendpunct}{\mcitedefaultseppunct}\relax
\EndOfBibitem
\bibitem[Andrzejak \latin{et~al.}(2019)Andrzejak, Sk{\'{o}}ra, and
  Petelenz]{Petelenz2019}
Andrzejak,~M.; Sk{\'{o}}ra,~T.; Petelenz,~P. {Is Vibrational Coherence a
  Byproduct of Singlet Exciton Fission?} \emph{The Journal of Physical
  Chemistry C} \textbf{2019}, \emph{123}, 91--101\relax
\mciteBstWouldAddEndPuncttrue
\mciteSetBstMidEndSepPunct{\mcitedefaultmidpunct}
{\mcitedefaultendpunct}{\mcitedefaultseppunct}\relax
\EndOfBibitem
\bibitem[Andrzejak \latin{et~al.}(2020)Andrzejak, Sk{\'{o}}ra, and
  Petelenz]{PetelenzJPCC2020}
Andrzejak,~M.; Sk{\'{o}}ra,~T.; Petelenz,~P. {Limitations of Generic
  Chromophore Concept for Femtosecond Vibrational Coherences}. \emph{The
  Journal of Physical Chemistry C} \textbf{2020}, \emph{124}, 3529--3535\relax
\mciteBstWouldAddEndPuncttrue
\mciteSetBstMidEndSepPunct{\mcitedefaultmidpunct}
{\mcitedefaultendpunct}{\mcitedefaultseppunct}\relax
\EndOfBibitem
\bibitem[Stern \latin{et~al.}(2017)Stern, Cheminal, Yost, Broch, Bayliss, Chen,
  Tabachnyk, Thorley, Greenham, Hodgkiss, Anthony, Head-Gordon, Musser, Rao,
  and Friend]{Stern2017}
Stern,~H.~L.; Cheminal,~A.; Yost,~S.~R.; Broch,~K.; Bayliss,~S.~L.; Chen,~K.;
  Tabachnyk,~M.; Thorley,~K.; Greenham,~N.; Hodgkiss,~J.~M.; Anthony,~J.;
  Head-Gordon,~M.; Musser,~A.~J.; Rao,~A.; Friend,~R.~H. {Vibronically coherent
  ultrafast triplet-pair formation and subsequent thermally activated
  dissociation control efficient endothermic singlet fission}. \emph{Nature
  Chemistry} \textbf{2017}, \emph{9}, 1205\relax
\mciteBstWouldAddEndPuncttrue
\mciteSetBstMidEndSepPunct{\mcitedefaultmidpunct}
{\mcitedefaultendpunct}{\mcitedefaultseppunct}\relax
\EndOfBibitem
\bibitem[Jean and Fleming(1995)Jean, and Fleming]{Jean1995}
Jean,~J.~M.; Fleming,~G.~R. {Competition between energy and phase relaxation in
  electronic curve crossing processes}. \emph{The Journal of Chemical Physics}
  \textbf{1995}, \emph{103}, 2092--2101\relax
\mciteBstWouldAddEndPuncttrue
\mciteSetBstMidEndSepPunct{\mcitedefaultmidpunct}
{\mcitedefaultendpunct}{\mcitedefaultseppunct}\relax
\EndOfBibitem
\bibitem[Paulus \latin{et~al.}(2020)Paulus, Adelman, Jamula, and
  McCusker]{Paulus2020}
Paulus,~B.~C.; Adelman,~S.~L.; Jamula,~L.; McCusker,~J. {Leveraging
  excited-state coherence for synthetic control of ultrafast dynamics}.
  \emph{Nature} \textbf{2020}, \emph{582}, 214--218\relax
\mciteBstWouldAddEndPuncttrue
\mciteSetBstMidEndSepPunct{\mcitedefaultmidpunct}
{\mcitedefaultendpunct}{\mcitedefaultseppunct}\relax
\EndOfBibitem
\bibitem[Delor \latin{et~al.}(2014)Delor, Scattergood, Sazanovich, Parker,
  Greetham, Meijer, Towrie, and Weinstein]{Weinstein2014}
Delor,~M.; Scattergood,~P.~A.; Sazanovich,~I.~V.; Parker,~A.~W.;
  Greetham,~G.~M.; Meijer,~A. J. H.~M.; Towrie,~M.; Weinstein,~J.~A. {Toward
  control of electron transfer in donor-acceptor molecules by bond-specific
  infrared excitation}. \emph{Science} \textbf{2014}, \emph{346},
  1492--1495\relax
\mciteBstWouldAddEndPuncttrue
\mciteSetBstMidEndSepPunct{\mcitedefaultmidpunct}
{\mcitedefaultendpunct}{\mcitedefaultseppunct}\relax
\EndOfBibitem
\bibitem[Scholes \latin{et~al.}(2017)Scholes, Fleming, Chen, Aspuru-Guzik,
  Buchleitner, Coker, Engel, van Grondelle, Ishizaki, Jonas, Lundeen, McCusker,
  Mukamel, Ogilvie, Olaya-Castro, Ratner, Spano, Whaley, and Zhu]{Scholes2017}
Scholes,~G.~D. \latin{et~al.}  {Using coherence to enhance function in chemical
  and biophysical systems}. \emph{Nature} \textbf{2017}, \emph{543},
  647--656\relax
\mciteBstWouldAddEndPuncttrue
\mciteSetBstMidEndSepPunct{\mcitedefaultmidpunct}
{\mcitedefaultendpunct}{\mcitedefaultseppunct}\relax
\EndOfBibitem
\bibitem[Witkowski and Moffitt(1960)Witkowski, and Moffitt]{Moffit1960}
Witkowski,~A.; Moffitt,~W. {Electronic Spectra of Dimers: Derivation of the
  Fundamental Vibronic Equation}. \emph{The Journal of Chemical Physics}
  \textbf{1960}, \emph{33}, 872--875\relax
\mciteBstWouldAddEndPuncttrue
\mciteSetBstMidEndSepPunct{\mcitedefaultmidpunct}
{\mcitedefaultendpunct}{\mcitedefaultseppunct}\relax
\EndOfBibitem
\bibitem[Fulton and Gouterman(1961)Fulton, and Gouterman]{Gouterman1961}
Fulton,~R.~L.; Gouterman,~M. {Vibronic Coupling. I. Mathematical Treatment for
  Two Electronic States}. \emph{The Journal of Chemical Physics} \textbf{1961},
  \emph{35}, 1059--1071\relax
\mciteBstWouldAddEndPuncttrue
\mciteSetBstMidEndSepPunct{\mcitedefaultmidpunct}
{\mcitedefaultendpunct}{\mcitedefaultseppunct}\relax
\EndOfBibitem
\bibitem[F{\"{o}}rster()]{Sinanoglu}
F{\"{o}}rster,~T. \emph{{Modern Quantum Chemistry, edited by O.
  Sinanogl$\backslash$u{\{}u{\}}}}; Academic Press Inc., New York. (1996),
  p~93\relax
\mciteBstWouldAddEndPuncttrue
\mciteSetBstMidEndSepPunct{\mcitedefaultmidpunct}
{\mcitedefaultendpunct}{\mcitedefaultseppunct}\relax
\EndOfBibitem
\bibitem[Berkelbach \latin{et~al.}(2014)Berkelbach, Hybertsen, and
  Reichman]{Berkelbach3}
Berkelbach,~T.~C.; Hybertsen,~M.~S.; Reichman,~D.~R. {Microscopic theory of
  singlet exciton fission. III. Crystalline pentacene}. \emph{The Journal of
  Chemical Physics} \textbf{2014}, \emph{141}, 74705\relax
\mciteBstWouldAddEndPuncttrue
\mciteSetBstMidEndSepPunct{\mcitedefaultmidpunct}
{\mcitedefaultendpunct}{\mcitedefaultseppunct}\relax
\EndOfBibitem
\bibitem[Phys \latin{et~al.}(2020)Phys, Accomasso, Granucci, and
  Wibowo]{Persico2020}
Phys,~J.~C.; Accomasso,~D.; Granucci,~G.; Wibowo,~M. {Delocalization effects in
  singlet fission : Comparing models with two and three interacting molecules
  Delocalization effects in singlet fission : Comparing models with two and
  three interacting molecules}. \textbf{2020}, \emph{244125}\relax
\mciteBstWouldAddEndPuncttrue
\mciteSetBstMidEndSepPunct{\mcitedefaultmidpunct}
{\mcitedefaultendpunct}{\mcitedefaultseppunct}\relax
\EndOfBibitem
\bibitem[Nagami \latin{et~al.}(2021)Nagami, Miyamoto, Sakai, and
  Nakano]{Nakano2021}
Nagami,~T.; Miyamoto,~H.; Sakai,~R.; Nakano,~M. {Stabilization of
  Charge-Transfer States in Pentacene Crystals and Its Role in Singlet
  Fission}. \emph{The Journal of Physical Chemistry C} \textbf{2021},
  \emph{125}, 2264--2275\relax
\mciteBstWouldAddEndPuncttrue
\mciteSetBstMidEndSepPunct{\mcitedefaultmidpunct}
{\mcitedefaultendpunct}{\mcitedefaultseppunct}\relax
\EndOfBibitem
\bibitem[Phys \latin{et~al.}(2020)Phys, Li, Parrish, Mart{\'{i}}nez, and
  Li]{Martinez2020}
Phys,~J.~C.; Li,~X.; Parrish,~R.~M.; Mart{\'{i}}nez,~T.~J.; Li,~X. {An ab
  initio exciton model for singlet fission An ab initio exciton model for
  singlet fission}. \textbf{2020}, \emph{184116}\relax
\mciteBstWouldAddEndPuncttrue
\mciteSetBstMidEndSepPunct{\mcitedefaultmidpunct}
{\mcitedefaultendpunct}{\mcitedefaultseppunct}\relax
\EndOfBibitem
\bibitem[Mirjani \latin{et~al.}(2014)Mirjani, Renaud, Gorczak, and
  Grozema]{Grozema2014}
Mirjani,~F.; Renaud,~N.; Gorczak,~N.; Grozema,~F.~C. {Theoretical Investigation
  of Singlet Fission in Molecular Dimers: The Role of Charge Transfer States
  and Quantum Interference}. \textbf{2014}, \relax
\mciteBstWouldAddEndPunctfalse
\mciteSetBstMidEndSepPunct{\mcitedefaultmidpunct}
{}{\mcitedefaultseppunct}\relax
\EndOfBibitem
\bibitem[Berkelbach \latin{et~al.}(2013)Berkelbach, Hybertsen, Reichman,
  Berkelbach, Hybertsen, and Reichman]{Berkelbach1}
Berkelbach,~T.~C.; Hybertsen,~M.~S.; Reichman,~D.~R.; Berkelbach,~T.~C.;
  Hybertsen,~M.~S.; Reichman,~D.~R. {Microscopic theory of singlet exciton
  fission . I . General formulation Microscopic theory of singlet exciton
  fission . I . General formulation}. \textbf{2013}, \emph{114102}\relax
\mciteBstWouldAddEndPuncttrue
\mciteSetBstMidEndSepPunct{\mcitedefaultmidpunct}
{\mcitedefaultendpunct}{\mcitedefaultseppunct}\relax
\EndOfBibitem
\bibitem[Hestand \latin{et~al.}(2015)Hestand, Yamagata, Xu, Sun, Zhong,
  Harutyunyan, Chen, Dai, Rao, and Spano]{Hestand2015}
Hestand,~N.~J.; Yamagata,~H.; Xu,~B.; Sun,~D.; Zhong,~Y.; Harutyunyan,~A.~R.;
  Chen,~G.; Dai,~H.-l.; Rao,~Y.; Spano,~F.~C. {Polarized Absorption in
  Crystalline Pentacene : Theory vs Experiment}. \emph{J. Chem. Phys.}
  \textbf{2015}, \emph{119}, 22137--22147\relax
\mciteBstWouldAddEndPuncttrue
\mciteSetBstMidEndSepPunct{\mcitedefaultmidpunct}
{\mcitedefaultendpunct}{\mcitedefaultseppunct}\relax
\EndOfBibitem
\bibitem[Coropceanu \latin{et~al.}(2007)Coropceanu, Demetrio, Filho, Olivier,
  Silbey, and Bre]{Bredas2007}
Coropceanu,~V.; Demetrio,~A.; Filho,~S.; Olivier,~Y.; Silbey,~R.; Bre,~J.-l.
  {Charge Transport in Organic Semiconductors}. \textbf{2007}, 926--952\relax
\mciteBstWouldAddEndPuncttrue
\mciteSetBstMidEndSepPunct{\mcitedefaultmidpunct}
{\mcitedefaultendpunct}{\mcitedefaultseppunct}\relax
\EndOfBibitem
\bibitem[Philpott(1969)]{Philpott1969}
Philpott,~M.~R. {Theory of the Vibrational Structure of Molecular Excitons.
  Soluble "One-Phonon” Models}. \emph{The Journal of Chemical Physics}
  \textbf{1969}, \emph{51}, 2616--2624\relax
\mciteBstWouldAddEndPuncttrue
\mciteSetBstMidEndSepPunct{\mcitedefaultmidpunct}
{\mcitedefaultendpunct}{\mcitedefaultseppunct}\relax
\EndOfBibitem
\bibitem[Roden \latin{et~al.}(2008)Roden, Eisfeld, and Briggs]{Briggs2008}
Roden,~J.; Eisfeld,~A.; Briggs,~J.~S. {The J- and H-bands of dye aggregate
  spectra: Analysis of the coherent exciton scattering (CES) approximation}.
  \emph{Chemical Physics} \textbf{2008}, \emph{352}, 258--266\relax
\mciteBstWouldAddEndPuncttrue
\mciteSetBstMidEndSepPunct{\mcitedefaultmidpunct}
{\mcitedefaultendpunct}{\mcitedefaultseppunct}\relax
\EndOfBibitem
\bibitem[Sahu \latin{et~al.}(2020)Sahu, Kurian, and Tiwari]{Sahu2020}
Sahu,~A.; Kurian,~J.~S.; Tiwari,~V. {Vibronic resonance is inadequately
  described by one-particle basis sets}. \emph{The Journal of Chemical Physics}
  \textbf{2020}, \emph{153}, 224114\relax
\mciteBstWouldAddEndPuncttrue
\mciteSetBstMidEndSepPunct{\mcitedefaultmidpunct}
{\mcitedefaultendpunct}{\mcitedefaultseppunct}\relax
\EndOfBibitem
\bibitem[Patra and Tiwari(2022)Patra, and Tiwari]{Patra2022}
Patra,~S.; Tiwari,~V. {Vibronic resonance along effective modes mediates
  selective energy transfer in excitonically coupled aggregates}. \emph{The
  Journal of Chemical Physics} \textbf{2022}, \emph{156}, 184115\relax
\mciteBstWouldAddEndPuncttrue
\mciteSetBstMidEndSepPunct{\mcitedefaultmidpunct}
{\mcitedefaultendpunct}{\mcitedefaultseppunct}\relax
\EndOfBibitem
\bibitem[Ba{\v{s}}inskaitė \latin{et~al.}(2014)Ba{\v{s}}inskaitė, Butkus,
  Abramavicius, and Valkunas]{Valkunas2014}
Ba{\v{s}}inskaitė,~E.; Butkus,~V.; Abramavicius,~D.; Valkunas,~L. {Vibronic
  models for nonlinear spectroscopy simulations}. \emph{Photosynthesis
  Research} \textbf{2014}, \emph{121}, 95--106\relax
\mciteBstWouldAddEndPuncttrue
\mciteSetBstMidEndSepPunct{\mcitedefaultmidpunct}
{\mcitedefaultendpunct}{\mcitedefaultseppunct}\relax
\EndOfBibitem
\bibitem[Filippini and Gramaccioli(1984)Filippini, and
  Gramaccioli]{Filipini1984}
Filippini,~G.; Gramaccioli,~C.~M. {Lattice-dynamical calculations for tetracene
  and pentacene}. \emph{Chemical Physics Letters} \textbf{1984}, \emph{104},
  50--53\relax
\mciteBstWouldAddEndPuncttrue
\mciteSetBstMidEndSepPunct{\mcitedefaultmidpunct}
{\mcitedefaultendpunct}{\mcitedefaultseppunct}\relax
\EndOfBibitem
\bibitem[Troisi and Orlandi(2006)Troisi, and Orlandi]{Troisi2006}
Troisi,~A.; Orlandi,~G. {Dynamics of the Intermolecular Transfer Integral in
  Crystalline Organic Semiconductors}. \emph{The Journal of Physical Chemistry
  A} \textbf{2006}, \emph{110}, 4065--4070\relax
\mciteBstWouldAddEndPuncttrue
\mciteSetBstMidEndSepPunct{\mcitedefaultmidpunct}
{\mcitedefaultendpunct}{\mcitedefaultseppunct}\relax
\EndOfBibitem
\bibitem[Dostál \latin{et~al.}(2014)Dostál, Mančal, Vácha, Pšenčík, and
  Zigmantas]{Mancal2014}
Dostál,~J.; Mančal,~T.; Vácha,~F.; Pšenčík,~J.; Zigmantas,~D. Unraveling
  the nature of coherent beatings in chlorosomes. \emph{The Journal of Chemical
  Physics} \textbf{2014}, \emph{140}, 115103\relax
\mciteBstWouldAddEndPuncttrue
\mciteSetBstMidEndSepPunct{\mcitedefaultmidpunct}
{\mcitedefaultendpunct}{\mcitedefaultseppunct}\relax
\EndOfBibitem
\bibitem[Seibt and Man{\v{c}}al(2018)Seibt, and Man{\v{c}}al]{Mancal2018}
Seibt,~J.; Man{\v{c}}al,~T. {Treatment of Herzberg-Teller and non-Condon
  effects in optical spectra with Hierarchical Equations of Motion}.
  \emph{Chemical Physics} \textbf{2018}, \emph{515}, 129--140\relax
\mciteBstWouldAddEndPuncttrue
\mciteSetBstMidEndSepPunct{\mcitedefaultmidpunct}
{\mcitedefaultendpunct}{\mcitedefaultseppunct}\relax
\EndOfBibitem
\bibitem[Ito \latin{et~al.}(2015)Ito, Nagami, and Nakano]{Nakano2015}
Ito,~S.; Nagami,~T.; Nakano,~M. {Density Analysis of Intra- and Intermolecular
  Vibronic Couplings toward Bath Engineering for Singlet Fission}. \emph{The
  Journal of Physical Chemistry Letters} \textbf{2015}, \emph{6},
  4972--4977\relax
\mciteBstWouldAddEndPuncttrue
\mciteSetBstMidEndSepPunct{\mcitedefaultmidpunct}
{\mcitedefaultendpunct}{\mcitedefaultseppunct}\relax
\EndOfBibitem
\bibitem[Cao \latin{et~al.}(2020)Cao, Cogdell, Coker, Duan, Hauer,
  Kleinekath{\"{o}}fer, Jansen, Man{\v{c}}al, Miller, Ogilvie, Prokhorenko,
  Renger, Tan, Tempelaar, Thorwart, Thyrhaug, Westenhoff, and
  Zigmantas]{Miller2020}
Cao,~J. \latin{et~al.}  {Quantum biology revisited}. \emph{Science Advances}
  \textbf{2020}, \emph{6}\relax
\mciteBstWouldAddEndPuncttrue
\mciteSetBstMidEndSepPunct{\mcitedefaultmidpunct}
{\mcitedefaultendpunct}{\mcitedefaultseppunct}\relax
\EndOfBibitem
\bibitem[Hestand and Spano(2018)Hestand, and Spano]{Hestand2018}
Hestand,~N.~J.; Spano,~F.~C. {Expanded Theory of H- and J-Molecular Aggregates:
  The Effects of Vibronic Coupling and Intermolecular Charge Transfer}.
  \emph{Chemical Reviews} \textbf{2018}, \emph{118}, 7069--7163\relax
\mciteBstWouldAddEndPuncttrue
\mciteSetBstMidEndSepPunct{\mcitedefaultmidpunct}
{\mcitedefaultendpunct}{\mcitedefaultseppunct}\relax
\EndOfBibitem
\bibitem[Sun \latin{et~al.}(2021)Sun, Liu, Hu, Zhang, Long, and Zhao]{Zhao2021}
Sun,~K.; Liu,~X.; Hu,~W.; Zhang,~M.; Long,~G.; Zhao,~Y. {Singlet fission
  dynamics and optical spectra of pentacene and its derivatives}. \emph{Phys.
  Chem. Chem. Phys.} \textbf{2021}, \emph{23}, 12654--12667\relax
\mciteBstWouldAddEndPuncttrue
\mciteSetBstMidEndSepPunct{\mcitedefaultmidpunct}
{\mcitedefaultendpunct}{\mcitedefaultseppunct}\relax
\EndOfBibitem
\bibitem[Hart \latin{et~al.}(2018)Hart, Silva, and Frontiera]{Frontiera2018}
Hart,~S.~M.; Silva,~W.~R.; Frontiera,~R.~R. {Femtosecond stimulated Raman
  evidence for charge-transfer character in pentacene singlet fission}.
  \emph{Chem. Sci.} \textbf{2018}, \emph{9}, 1242--1250\relax
\mciteBstWouldAddEndPuncttrue
\mciteSetBstMidEndSepPunct{\mcitedefaultmidpunct}
{\mcitedefaultendpunct}{\mcitedefaultseppunct}\relax
\EndOfBibitem
\bibitem[Castellanos and Huo(2017)Castellanos, and Huo]{Huo2017}
Castellanos,~M.~A.; Huo,~P. {Enhancing Singlet Fission Dynamics by Suppressing
  Destructive Interference between Charge-Transfer Pathways}. \emph{The Journal
  of Physical Chemistry Letters} \textbf{2017}, \emph{8}, 2480--2488\relax
\mciteBstWouldAddEndPuncttrue
\mciteSetBstMidEndSepPunct{\mcitedefaultmidpunct}
{\mcitedefaultendpunct}{\mcitedefaultseppunct}\relax
\EndOfBibitem
\bibitem[Montoya-Castillo \latin{et~al.}(2015)Montoya-Castillo, Berkelbach, and
  Reichman]{Castillo2015}
Montoya-Castillo,~A.; Berkelbach,~T.~C.; Reichman,~D.~R. {Extending the
  applicability of Redfield theories into highly non-Markovian regimes}.
  \emph{The Journal of Chemical Physics} \textbf{2015}, \emph{143},
  194108\relax
\mciteBstWouldAddEndPuncttrue
\mciteSetBstMidEndSepPunct{\mcitedefaultmidpunct}
{\mcitedefaultendpunct}{\mcitedefaultseppunct}\relax
\EndOfBibitem
\bibitem[Mukamel(1990)]{Mukamel1990}
Mukamel,~S. Femtosecond Optical Spectroscopy: A Direct Look at Elementary
  Chemical Events. \emph{Annual Review of Physical Chemistry} \textbf{1990},
  \emph{41}, 647--681\relax
\mciteBstWouldAddEndPuncttrue
\mciteSetBstMidEndSepPunct{\mcitedefaultmidpunct}
{\mcitedefaultendpunct}{\mcitedefaultseppunct}\relax
\EndOfBibitem
\bibitem[Higgins \latin{et~al.}(2021)Higgins, Lloyd, Sohail, Allodi, Otto,
  Saer, Wood, Massey, Ting, Blankenship, and Engel]{Engel2021}
Higgins,~J.~S.; Lloyd,~L.~T.; Sohail,~S.~H.; Allodi,~M.~A.; Otto,~J.~P.;
  Saer,~R.~G.; Wood,~R.~E.; Massey,~S.~C.; Ting,~P.-C.; Blankenship,~R.~E.;
  Engel,~G.~S. {Photosynthesis tunes quantum-mechanical mixing of electronic
  and vibrational states to steer exciton energy transfer}. \emph{Proceedings
  of the National Academy of Sciences} \textbf{2021}, \emph{118},
  e2018240118\relax
\mciteBstWouldAddEndPuncttrue
\mciteSetBstMidEndSepPunct{\mcitedefaultmidpunct}
{\mcitedefaultendpunct}{\mcitedefaultseppunct}\relax
\EndOfBibitem
\bibitem[Yarkony(1996)]{Yarkony1996}
Yarkony,~D.~R. {Diabolical conical intersections}. \emph{Reviews of Modern
  Physics} \textbf{1996}, \emph{68}, 985--1013\relax
\mciteBstWouldAddEndPuncttrue
\mciteSetBstMidEndSepPunct{\mcitedefaultmidpunct}
{\mcitedefaultendpunct}{\mcitedefaultseppunct}\relax
\EndOfBibitem
\bibitem[Farrow \latin{et~al.}(2008)Farrow, Qian, Smith, Ferro, and
  Jonas]{Jonas2008}
Farrow,~D.~A.; Qian,~W.; Smith,~E.~R.; Ferro,~A.~A.; Jonas,~D.~M. {Polarized
  pump-probe measurements of electronic motion via a conical intersection}.
  \emph{The Journal of Chemical Physics} \textbf{2008}, \emph{128},
  144510\relax
\mciteBstWouldAddEndPuncttrue
\mciteSetBstMidEndSepPunct{\mcitedefaultmidpunct}
{\mcitedefaultendpunct}{\mcitedefaultseppunct}\relax
\EndOfBibitem
\bibitem[Farrow \latin{et~al.}(2008)Farrow, Smith, Qian, and Jonas]{Jonas2008b}
Farrow,~D.~A.; Smith,~E.~R.; Qian,~W.; Jonas,~D.~M. {The polarization
  anisotropy of vibrational quantum beats in resonant pump- probe experiments :
  Diagrammatic calculations for square symmetric molecules The polarization
  anisotropy of vibrational quantum beats in resonant pump-probe experiments :
  Diagramma}. \textbf{2008}, \emph{174509}\relax
\mciteBstWouldAddEndPuncttrue
\mciteSetBstMidEndSepPunct{\mcitedefaultmidpunct}
{\mcitedefaultendpunct}{\mcitedefaultseppunct}\relax
\EndOfBibitem
\bibitem[Kitney-Hayes \latin{et~al.}(2014)Kitney-Hayes, Ferro, Tiwari, and
  Jonas]{KitneyHayes2014}
Kitney-Hayes,~K.~A.; Ferro,~A.~A.; Tiwari,~V.; Jonas,~D.~M. {Two-dimensional
  Fourier transform electronic spectroscopy at a conical intersection}.
  \emph{The Journal of Chemical Physics} \textbf{2014}, \emph{140},
  124312\relax
\mciteBstWouldAddEndPuncttrue
\mciteSetBstMidEndSepPunct{\mcitedefaultmidpunct}
{\mcitedefaultendpunct}{\mcitedefaultseppunct}\relax
\EndOfBibitem
\bibitem[Musser \latin{et~al.}(2015)Musser, Liebel, Schnedermann, Wende, Kehoe,
  Rao, and Kukura]{Musser2015}
Musser,~A.~J.; Liebel,~M.; Schnedermann,~C.; Wende,~T.; Kehoe,~T.~B.; Rao,~A.;
  Kukura,~P. {Evidence for conical intersection dynamics mediating ultrafast
  singlet exciton fission}. \emph{Nature Physics} \textbf{2015}, \emph{11},
  352\relax
\mciteBstWouldAddEndPuncttrue
\mciteSetBstMidEndSepPunct{\mcitedefaultmidpunct}
{\mcitedefaultendpunct}{\mcitedefaultseppunct}\relax
\EndOfBibitem
\bibitem[Wilson \latin{et~al.}(2011)Wilson, Rao, Clark, Kumar, Brida, Cerullo,
  and Friend]{Friend2011}
Wilson,~M. W.~B.; Rao,~A.; Clark,~J.; Kumar,~R. S.~S.; Brida,~D.; Cerullo,~G.;
  Friend,~R.~H. {Ultrafast Dynamics of Exciton Fission in Polycrystalline
  Pentacene}. \textbf{2011}, 11830--11833\relax
\mciteBstWouldAddEndPuncttrue
\mciteSetBstMidEndSepPunct{\mcitedefaultmidpunct}
{\mcitedefaultendpunct}{\mcitedefaultseppunct}\relax
\EndOfBibitem
\bibitem[Kim \latin{et~al.}(2020)Kim, Kim, Burger, Park, Kling, Kim, and
  Joo]{Joo2020}
Kim,~J.; Kim,~C.~H.; Burger,~C.; Park,~M.; Kling,~M.~F.; Kim,~D.~E.; Joo,~T.
  {Non-Born–Oppenheimer Molecular Dynamics Observed by Coherent Nuclear Wave
  Packets}. \emph{The Journal of Physical Chemistry Letters} \textbf{2020},
  \emph{11}, 755--761\relax
\mciteBstWouldAddEndPuncttrue
\mciteSetBstMidEndSepPunct{\mcitedefaultmidpunct}
{\mcitedefaultendpunct}{\mcitedefaultseppunct}\relax
\EndOfBibitem
\bibitem[{Thorsm{\o}lle, X.} \latin{et~al.}(2009){Thorsm{\o}lle, X.}, Averitt,
  Demsar, Smith, Tretiak, Martin, Chi, Crone, Ramirez, and Taylor]{Taylor2009}
{Thorsm{\o}lle, X.},~V.~K.; Averitt,~R.~D.; Demsar,~J.; Smith,~D.~L.;
  Tretiak,~S.; Martin,~R.~L.; Chi,~X.; Crone,~B.~K.; Ramirez,~A.~P.;
  Taylor,~A.~J. {Morphology Effectively Controls Singlet-Triplet Exciton
  Relaxation and Charge Transport in Organic Semiconductors}. \emph{Phys. Rev.
  Lett.} \textbf{2009}, \emph{102}, 17401\relax
\mciteBstWouldAddEndPuncttrue
\mciteSetBstMidEndSepPunct{\mcitedefaultmidpunct}
{\mcitedefaultendpunct}{\mcitedefaultseppunct}\relax
\EndOfBibitem
\bibitem[Jonas(2003)]{JonasARPC2003}
Jonas,~D.~M. {Two-Dimensional Femtosecond Spectroscopy}. \emph{Annu. Rev. Phys.
  Chem.} \textbf{2003}, \emph{54}, 425--463\relax
\mciteBstWouldAddEndPuncttrue
\mciteSetBstMidEndSepPunct{\mcitedefaultmidpunct}
{\mcitedefaultendpunct}{\mcitedefaultseppunct}\relax
\EndOfBibitem
\bibitem[Qian and Jonas(2003)Qian, and Jonas]{Jonas2003}
Qian,~W.; Jonas,~D.~M. {Role of cyclic sets of transition dipoles in the
  pump–probe polarization anisotropy: Application to square symmetric
  molecules and perpendicular chromophore pairs}. \emph{The Journal of Chemical
  Physics} \textbf{2003}, \emph{119}, 1611--1622\relax
\mciteBstWouldAddEndPuncttrue
\mciteSetBstMidEndSepPunct{\mcitedefaultmidpunct}
{\mcitedefaultendpunct}{\mcitedefaultseppunct}\relax
\EndOfBibitem
\bibitem[Fuller and Ogilvie(2015)Fuller, and Ogilvie]{Fuller2015}
Fuller,~F.~D.; Ogilvie,~J.~P. {Experimental Implementations of Two-Dimensional
  Fourier Transform Electronic Spectroscopy}. \emph{Annual Review of Physical
  Chemistry} \textbf{2015}, \emph{66}, 667--690\relax
\mciteBstWouldAddEndPuncttrue
\mciteSetBstMidEndSepPunct{\mcitedefaultmidpunct}
{\mcitedefaultendpunct}{\mcitedefaultseppunct}\relax
\EndOfBibitem
\bibitem[Zanni \latin{et~al.}(2001)Zanni, Ge, Kim, and Hochstrasser]{Zanni2001}
Zanni,~M.~T.; Ge,~N.-H.; Kim,~Y.~S.; Hochstrasser,~R.~M. {Two-dimensional IR
  spectroscopy can be designed to eliminate the diagonal peaks and expose only
  the crosspeaks needed for structure determination}. \emph{Proceedings of the
  National Academy of Sciences} \textbf{2001}, \emph{98}, 11265--11270\relax
\mciteBstWouldAddEndPuncttrue
\mciteSetBstMidEndSepPunct{\mcitedefaultmidpunct}
{\mcitedefaultendpunct}{\mcitedefaultseppunct}\relax
\EndOfBibitem
\bibitem[Thyrhaug \latin{et~al.}(2018)Thyrhaug, Tempelaar, Alcocer,
  {\v{Z}}{\'{i}}dek, B{\'{i}}na, Knoester, Jansen, and Zigmantas]{Thyrhaug2018}
Thyrhaug,~E.; Tempelaar,~R.; Alcocer,~M. J.~P.; {\v{Z}}{\'{i}}dek,~K.;
  B{\'{i}}na,~D.; Knoester,~J.; Jansen,~T. L.~C.; Zigmantas,~D. {Identification
  and characterization of diverse coherences in the Fenna–Matthews–Olson
  complex}. \emph{Nature Chemistry} \textbf{2018}, \emph{10}, 780--786\relax
\mciteBstWouldAddEndPuncttrue
\mciteSetBstMidEndSepPunct{\mcitedefaultmidpunct}
{\mcitedefaultendpunct}{\mcitedefaultseppunct}\relax
\EndOfBibitem
\bibitem[Farrell \latin{et~al.}(2022)Farrell, Yang, and Zanni]{Zanni2022}
Farrell,~K.~M.; Yang,~N.; Zanni,~M.~T. {A polarization scheme that resolves
  cross-peaks with transient absorption and eliminates diagonal peaks in 2D
  spectroscopy}. \emph{Proceedings of the National Academy of Sciences}
  \textbf{2022}, \emph{119}, e2117398119\relax
\mciteBstWouldAddEndPuncttrue
\mciteSetBstMidEndSepPunct{\mcitedefaultmidpunct}
{\mcitedefaultendpunct}{\mcitedefaultseppunct}\relax
\EndOfBibitem
\bibitem[Pensack \latin{et~al.}(2015)Pensack, Tilley, Parkin, Lee, Payne, Gao,
  Jahnke, Oblinsky, Li, Anthony, Seferos, and Scholes]{Pensack2015}
Pensack,~R.~D.; Tilley,~A.~J.; Parkin,~S.~R.; Lee,~T.~S.; Payne,~M.~M.;
  Gao,~D.; Jahnke,~A.~A.; Oblinsky,~D.~G.; Li,~P.-F.; Anthony,~J.~E.;
  Seferos,~D.~S.; Scholes,~G.~D. Exciton Delocalization Drives Rapid Singlet
  Fission in Nanoparticles of Acene Derivatives. \emph{Journal of the American
  Chemical Society} \textbf{2015}, \emph{137}, 6790--6803, PMID: 25946670\relax
\mciteBstWouldAddEndPuncttrue
\mciteSetBstMidEndSepPunct{\mcitedefaultmidpunct}
{\mcitedefaultendpunct}{\mcitedefaultseppunct}\relax
\EndOfBibitem
\bibitem[Krishnapriya \latin{et~al.}(2019)Krishnapriya, Musser, and
  Patil]{Patil2019}
Krishnapriya,~K.~C.; Musser,~A.~J.; Patil,~S. {Molecular Design Strategies for
  Efficient Intramolecular Singlet Exciton Fission}. \emph{ACS Energy Letters}
  \textbf{2019}, \emph{4}, 192--202\relax
\mciteBstWouldAddEndPuncttrue
\mciteSetBstMidEndSepPunct{\mcitedefaultmidpunct}
{\mcitedefaultendpunct}{\mcitedefaultseppunct}\relax
\EndOfBibitem
\end{mcitethebibliography}

\end{document}